\documentclass[twocolumn]{aastex6}
\usepackage{graphicx}
\usepackage{natbib}
\bibpunct{(}{)}{;}{a}{}{,} 
\usepackage[varg]{txfonts}
\usepackage{hyperref}

\usepackage{color}

\newcommand{\snr} {\mbox{S/N}}

\newcommand{\ie}{i.\,e.}

\newcommand{\eg}{e.\,g.}

\newcommand{\teff}{$T_{{\rm eff}}$}
\newcommand{\kms}{\mbox{km\,s$^{-1}$}}
\newcommand{\ms}{\mbox{m s$^{-1}$}}
\newcommand{\gcm}{\mbox{g cm$^{-3}$}}

\newcommand{\vmicro} {\mbox{$\xi_{\rm t}$}}
\newcommand{\gfeh} {\mbox{$[{\rm Fe}/{\rm H}]$}}

\newcommand{\mearth}{\mbox{M$_\oplus$}}
\newcommand{\rearth}{\mbox{R$_\oplus$}}
\newcommand{\roearth}{\mbox{$\rho_\oplus$}}
\newcommand{\fearth}{\mbox{$F_\oplus$}}
\newcommand{\msun}{\mbox{M$_\odot$}}
\newcommand{\rsun}{\mbox{R$_\odot$}}
\newcommand{\mstar}{\mbox{M$_\star$}}
\newcommand{\rstar}{\mbox{R$_\star$}}
\newcommand{\mjup}{\mbox{M$_{\rm jup}$}}
\newcommand{\logRHK}{\mbox{$\log {\rm R}^{\prime}_{\rm HK}$}}

\newcommand{\CCFa}{\mbox{CCF$_{\rm A}$}}
\newcommand{\CCFb}{\mbox{CCF$_{\rm B}$}}
\newcommand{\Kepb}{\mbox{{\it Kepler}-19b}}
\newcommand{\Kepc}{\mbox{{\it Kepler}-19c}}
\newcommand{\Kepd}{\mbox{{\it Kepler}-19d}}
\newcommand{\Kb}{\mbox{{\it K}19b}}
\newcommand{\Kc}{\mbox{{\it K}19c}}
\newcommand{\Kd}{\mbox{{\it K}19d}}
\newcommand{\Ks}{\mbox{{\it Kepler}-19}}

\begin{document}

\title{The Kepler-19 system: a thick-envelope super-Earth with two
    Neptune-mass companions characterized using Radial Velocities and Transit Timing Variations}

\author{
Luca Malavolta\altaffilmark{1,2},
Luca Borsato\altaffilmark{1,2},
Valentina Granata\altaffilmark{1,2},
Giampaolo Piotto\altaffilmark{1,2},
Eric Lopez\altaffilmark{3},
Andrew Vanderburg\altaffilmark{4},
Pedro Figueira\altaffilmark{5},
Annelies Mortier\altaffilmark{6},
Valerio Nascimbeni\altaffilmark{1,2},
Laura Affer\altaffilmark{7},
Aldo S. Bonomo\altaffilmark{8},
Francois Bouchy\altaffilmark{9},
Lars~A.~Buchhave\altaffilmark{10},
David Charbonneau\altaffilmark{4},
Andrew Collier Cameron\altaffilmark{6},
Rosario Cosentino\altaffilmark{11},
Courtney D. Dressing\altaffilmark{12,13},
Xavier Dumusque\altaffilmark{9},
Aldo F. M. Fiorenzano\altaffilmark{11},
Avet Harutyunyan\altaffilmark{11},
Rapha\"elle D. Haywood\altaffilmark{4},
John Asher Johnson\altaffilmark{4},
David W. Latham\altaffilmark{4},
Mercedes Lopez-Morales\altaffilmark{4},
Christophe Lovis\altaffilmark{9},
Michel Mayor\altaffilmark{9},
Giusi Micela\altaffilmark{7},
Emilio Molinari\altaffilmark{11,14},
Fatemeh Motalebi\altaffilmark{9},
Francesco Pepe\altaffilmark{9},
David F. Phillips\altaffilmark{4},
Don Pollacco\altaffilmark{15},
Didier Queloz\altaffilmark{9,16},
Ken Rice\altaffilmark{3},
Dimitar Sasselov\altaffilmark{4},
Damien S\'egransan\altaffilmark{9},
Alessandro Sozzetti\altaffilmark{8},
St\'ephane Udry\altaffilmark{9},
Chris Watson\altaffilmark{17}
}

\altaffiltext{1}{Dipartimento di Fisica e Astronomia ``Galileo Galilei", Universita'di Padova, Vicolo dell'Osservatorio 3, 35122 Padova, Italy; \href{mailto:luca.malavolta@unipd,it}{luca.malavolta@unipd.it}}
\altaffiltext{2}{INAF - Osservatorio Astronomico di Padova, Vicolo dell'Osservatorio 5, 35122 Padova, Italy}
\altaffiltext{3}{SUPA, Institute for Astronomy, University of Edinburgh, Royal Observatory, Blackford Hill, Edinburgh, EH93HJ, UK}
\altaffiltext{4}{Harvard-Smithsonian Center for Astrophysics, 60 Garden Street, Cambridge, Massachusetts 02138, USA}
\altaffiltext{5}{Instituto de Astrof\' isica e Ci\^encias do Espa\c{c}o, Universidade do Porto, CAUP, Rua das Estrelas, PT4150-762 Porto, Portugal}
\altaffiltext{6}{Centre for Exoplanet Science, SUPA, School of Physics and Astronomy, University of St. Andrews, St. Andrews KY16 9SS, UK}
\altaffiltext{7}{INAF - Osservatorio Astronomico di Palermo, Piazza del Parlamento 1, 90124 Palermo, Italy}
\altaffiltext{8}{INAF - Osservatorio Astrofisico di Torino, via Osservatorio 20, 10025 Pino Torinese, Italy}
\altaffiltext{9}{Observatoire Astronomique de l'Universit\'e de Gen\`eve, 51 ch. des Maillettes, 1290 Versoix, Switzerland}
\altaffiltext{10}{Centre for Star and Planet Formation, Natural History Museum of Denmark \& Niels Bohr Institute, University of Copenhagen, \O ster Voldgade 5-7, DK-1350 Copenhagen K, Denmark}
\altaffiltext{11}{INAF - Fundaci\'on Galileo Galilei, Rambla Jos\'e Ana Fernandez P\'erez 7, 38712 Bre\~na Baja, Spain}
\altaffiltext{12}{Division of Geological and Planetary Sciences, California Institute of Technology, Pasadena, CA 91125, USA}
\altaffiltext{13}{NASA Sagan Fellow}
\altaffiltext{14}{INAF - IASF Milano, via Bassini 15, 20133, Milano, Italy}
\altaffiltext{15}{Department of Physics, University of Warwick, Gibbet Hill Road, Coventry CV4 7AL, UK}
\altaffiltext{16}{Cavendish Laboratory, J J Thomson Avenue, Cambridge CB3 0HE, UK}
\altaffiltext{17}{Astrophysics Research Centre, School of Mathematics and Physics, Queen’s University Belfast, Belfast BT7 1NN, UK}

\begin{abstract}
We report a detailed characterization of the \Ks\ system. This star was previously known to host a transiting planet with a period of 9.29 days, a radius of 2.2 \rearth\ and an upper limit on the mass of 20 \mearth. The presence of a second, non-transiting planet was inferred from the transit time variations (TTVs) of \Kepb\, over 8 quarters of \textit{Kepler} photometry, although neither mass nor period could be determined. By combining new TTVs measurements from all the \textit{Kepler} quarters and  91 high-precision radial velocities obtained with the HARPS-N spectrograph, we measured through dynamical simulations a mass of $8.4 \pm 1.6$ \mearth\ for \Kepb. From the same data, assuming system coplanarity, we determined an orbital period of 28.7 days and a  mass of $13.1 \pm 2.7$ \mearth\ for \Kepc\ and discovered a Neptune-like planet with a mass of $20.3 \pm 3.4$ \mearth\ on a 63 days orbit.
By comparing dynamical simulations with non-interacting Keplerian orbits, we concluded that neglecting interactions between planets may lead to systematic errors that could hamper the precision in the orbital parameters when the dataset spans several years.

With a density of $4.32 \pm 0.87$ \gcm\ ($0.78 \pm 0.16$ \roearth) \Kepb\ belongs to the group of planets with a rocky core and a significant fraction of volatiles, in opposition to low-density planets characterized by transit-time variations only and the increasing number of rocky planets with Earth-like density. \Ks\ joins the small number of systems that reconcile transit timing variation and radial velocity measurements.
\end{abstract}

\shorttitle{The \Ks\ system}
\shortauthors{Malavolta et al.}

\maketitle

\section{Introduction}\label{sec:introduction}

After the discovery of thousands of planets with radii smaller than 2.7 \rearth\ from the NASA {\it Kepler} mission \citep{Borucki2011,Coughlin2016}, a consistent effort has been devoted to understanding the formation scenario and chemical composition of such planets (\eg , \citealt{Weiss2014}, \citealt{Dressing2015} and \citealt{WolfgangLopez2015}). To distinguish between a rocky composition and the presence of a thick envelope of water or volatile elements, the radius derived from the transit depth must be coupled with a precise mass determination (better than 20\%), either from radial velocity (RV) measurements or transit timing variations (TTVs). Planets that have been characterized with such level of precision appear to fall into two populations, a first one following an Earth-like composition and a second one with planets larger than 2 Earth radii requiring a significant fraction of volatiles (\eg , \citealt{Rogers2015}, \citealt{Gettel2016} and \citealt{LopezMorales2016}). Recently, improved mass and radius determinations of known planets have uncovered the existence of super-Earths which fall between these two populations, such as 55 Cancri e \citep{Demory2016} and {\it Kepler}-20b \citep{Buchhave2016}. 

We need more planets in that mass regime with precise mass and radius measurements to understand the undergoing physics.
For this reason we carried out a RV follow-up of \Kepb\ (hereafter \Kb ), a planet with period of $9.287$ days and radius $2.209 \pm 0.048 $ \rearth , orbiting a relatively bright ($V=12.1,  K=11.9$) solar-type star (\teff\ $= 5541 \pm 60$ K, $\log g = 4.59 \pm 0.10$, \gfeh\ $ = -0.13 \pm 0.06$). The planet was detected by \cite{Borucki2011} and subsequently validated by \citet[hereafter B11]{Ballard2011} using adaptive optics and speckle imaging to exclude a secondary source in the {\it Kepler} light curve, {\it Spitzer} observations to verify the achromaticity of the transit and Keck-HIRES high-resolution spectroscopy to rule out the presence of massive, non-planetary perturbers. From {\tt BLENDER} analysis \citep{Torres2011} the probability of a false-positive scenario was constrained to less than $1.5 \times 10^{-4}$.
  RVs measured on high-resolution spectra were consistent with a mass of 1.5 \mearth\ (0.5 \ms ) and activity-induced RV jitter of 4 \ms, but ultimately they lacked the required precision for a robust determination of the planetary mass, and only an upper limit of 20.3 \mearth\ was set. The existence of an additional planet, \Kepc\ (hereafter \Kc ), with period $\lesssim 160 $ days and mass $\lesssim 6$ \mjup , and a further confirmation of the planetary nature of \Kb\  were inferred by B11 from the presence of TTVs on 8 quarters of \Kb\ light curve.

  In this paper we couple high-precision RV measurements obtained with HARPS-N with updated measurements of transit times ($T_0$) encompassing all 17 quarters of {\it Kepler} data to determine the orbital parameters of \Kb , \Kc\ and a third, previously unknown planet in the system, \Kepd\ (hereafter \Kd ).
TTV and RV datasets are analyzed independently, to understand which constraints they can provide to the characterization of the system. Subsequently, a simultaneous TTV and RV fit is performed using dynamical simulation to take into account gravitational interactions between planets. We perform this analysis under the assumption of coplanarity between planets, and then investigate the effect of different mutual inclinations on the goodness of the fit. We confirm that only the inner planet is seen transiting the host star.
A comparison between RV obtained from dynamical simulations and when assuming non-interacting planets is performed.
We conclude describing the role of \Kb\ in understanding the bulk densities of small planets.

\section{Radial Velocities}\label{sec:radial_velocities}

We collected 101 spectra using HARPS-N at the Telescopio Nazionale Galileo, in La Palma. Observations spanned over two years, from June 2012 to November 2014, overlapping the {\it Kepler} observations during the first year. Every observation consisted of a 30 minute exposure, with a median signal-to-noise ratio of 37 at 550 nm, corresponding to a RV nominal error of 2.8 \ms . Given the faintness of the target, observations were gathered with the {\it objAB} setting, \ie , the second fiber (fiber B) was observing the sky instead of acquiring a simultaneous thorium-argon (ThAr) lamp spectrum. Several observations demonstrated that the stability of the instrument over 24 hours is within 1 \ms (\eg , \citealt{Cosentino2014}), thus the precision of the measurements was dictated largely by photon noise.
Data were reduced using the standard Data Reduction Software (DRS) using a G8 flux template (the closest available one to the spectral type of the target) to correct for variations in the flux distribution as a function of the wavelength, and a G2 binary mask was used to perform the cross-correlation \citep{Baranne1996, Pepe2002}. The resulting RV data with their formal 1-$\sigma$ uncertainties, the Full Width Half Maximum (FWHM) of the cross-correlation function (CCF) and its contrast (\ie , the depth normalized to the continuum), the bisector inverse span (BIS), and the \logRHK\ activity index  are listed in Table
\ref{table:RV_table}.

\begin{table*}
  \caption{HARPS-N radial velocities and ancillary measurements of Kepler-19. Epoch of the observations, RV with associated noise, FWHM and contrast of the CCF, inverse bisector span,\logRHK\ activity index with the associated error, and a flag indicating if the data has been contaminated by the Moon (1) or not (0), see Section~\ref{sec:effect-moon-illum}.}
\label{table:RV_table}      
\centering                                      
\begin{tabular}{ccccccccc}          
\tableline\tableline                        
  BJD$_{\rm UTC}$ & RV & $\sigma _{RV}$ & FWHM & Contrast & BIS & \logRHK & $\sigma_{\log {\rm R}^{\prime}_{\rm HK} }$ & Moon flag \\
 $[$d$]$  & [\ms ] & [\ms ] & [\kms] & &   [\ms ] & [dex] & [dex] & \\

 \tableline                                 
2456100.608 & -10608.31 & 2.31 &  6.745 & 48.83 & -41.90 & -4.975 & 0.039 & 0 \\
2456100.629 & -10610.99 & 2.09 &  6.735 & 48.91 & -40.73 & -5.039 & 0.031 & 0 \\
2456101.606 & -10613.07 & 3.56 &  6.732 & 48.70 & -55.78 & -4.961 & 0.060 & 0 \\
... &   ...  & ... &   ...  &  ...   &   ... & ... & ... & . \\

\tableline                        
\tableline
\end{tabular}
  \tablecomments{Table \ref{table:RV_table} is published in its entirety in the machine-readable format. A portion is shown here for guidance regarding its form and content.}
\end{table*}

\subsection{Effect of Moon Illumination}\label{sec:effect-moon-illum}
A simple procedure was adopted to check the influence of the moon illumination on the science fiber (labeled as fiber A). First, the cross-correlation function of the sky spectrum acquired with fiber B, \CCFb , was recomputed\footnote{When using the objAB setting, \CCFb\ is computed by the DRS without flux correction by default.} using the same flux correction coefficients as for the target (\CCFa ) for that specific acquisition. Then, \CCFb\ was subtracted to the corresponding \CCFa\ and radial velocities were computed again using the script from \cite{Figueira2013}, which uses the same algorithm implemented in the DRS. For 10 observations the difference between the sky-corrected RV and the DRS RV was greater than twice the photon noise, so we rejected those observations, and used the remaining 91 RVs from the DRS in the following analysis. A flag has been included in Table~\ref{table:RV_table} to identify the rejected observations.

While the rejected observations have in common a fraction of the illuminated Moon greater than 0.9 and a barycentric RV correction within 15 \kms\ from the absolute RV of the target star, not all the observations that met this criterium were affected by the sky contamination, suggesting that other unidentified factors can determine whether or not contamination is negligible. An in-depth analysis of the outcome of the observations is then advised when measuring RVs for faint stars.

\section{{\it Kepler} photometry}\label{sec:kepler-photometry}
\Ks\ was initially observed in long-cadence (LC) mode during the first 0-2 quarters, and then in short-cadence (SC) mode from quarter 3 until the end of the mission in 2013 (at quarter 17). At the time of publishing, B11 had at their disposal only the first 8 quarters.
In an effort to make use of any additional information coming from {\it Kepler} photometry, we redetermined the transit times for all the quarters now available, in both LC and SC light curves. Quarters already analyzed by B11 were examined as well, to validate our $T_0$ determination and to provide a homogeneous set of measurements.

Transit identification was performed by propagating the linear ephemeris of B11, with the inclusion of 3 hours of pre-ingress and post-egress around the expected transit time.

For each transit time, we firstly detrended the transit light curve with a polynomial between $1^\textrm{st}$ and $10^\textrm{th}$ degree, with the best-fit degree chosen according to the Bayesian information criterion (BIC).
Then, we determined a new $T_0$ guess with an automatic selection among different search algorithms
\footnote{Levenberg-Marquardt \citep{More1980}, Nelder-Mead  \citep{Nelder01011965, Wright96a}, COBYLA \citep{Powell1994}  as implemented in {\tt scipy.optimize}, and the Affine Invariant  Markov Chain Monte Carlo (MCMC) Ensemble sampler implemented in  {\tt emcee} \citep{ForemanMackey2013} available at \url{https://github.com/dfm/emcee}} fitting the \citet{MA2002ApJ} transit model, implemented in {\tt PyTransit}\footnote{Available at \url{https://github.com/hpparvi/PyTransit}} \citep{Parviainen2015}, fixing all other parameters to the literature value. Finally, the $T_0$s were refined using {\tt JKTEBOP} program \citep{JKTEBOP2004} and the associated errors were determined with a classical bootstrap approach.

\begin{figure}[htbp]
\includegraphics[width=\linewidth]{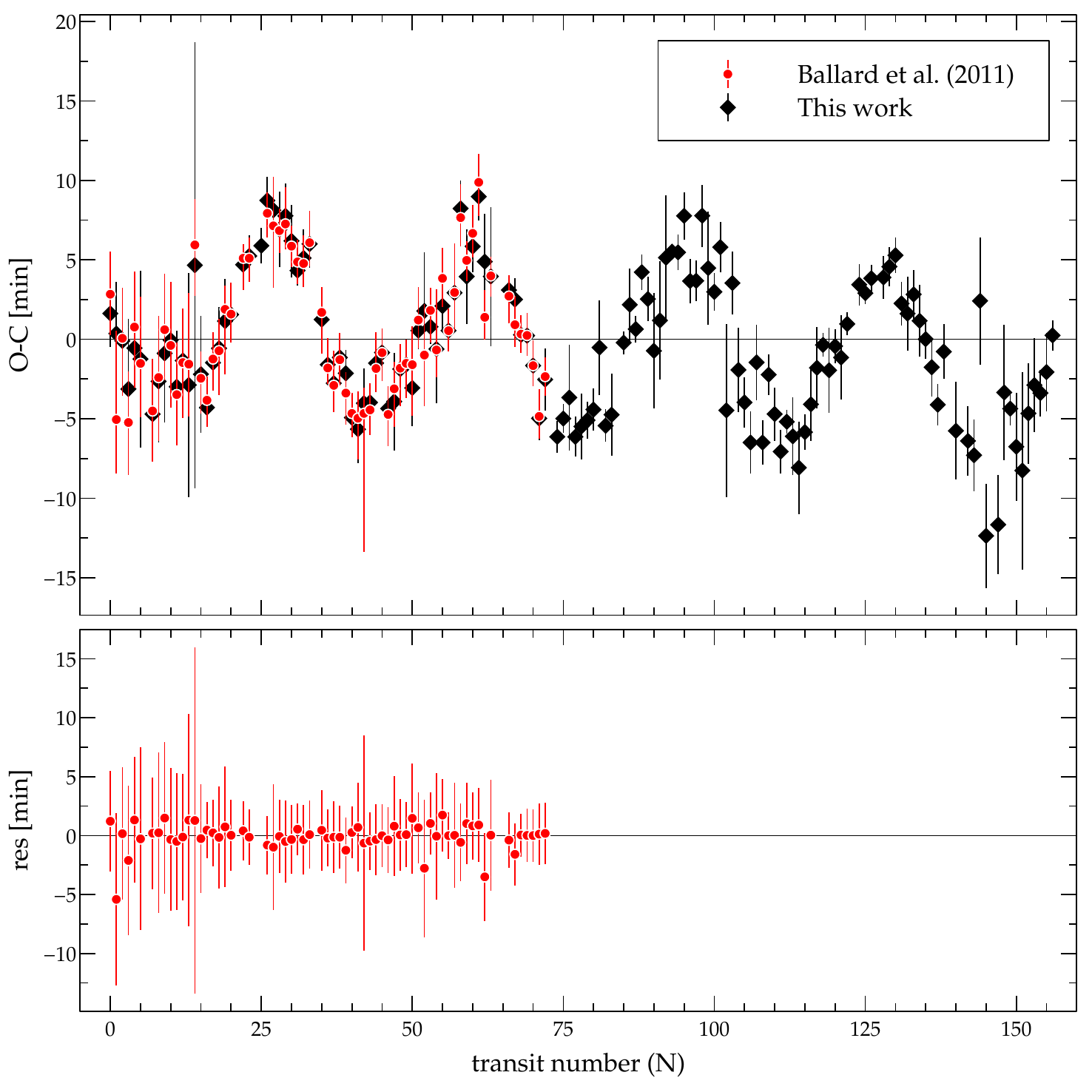}
\caption{The difference between the observed and the predicted (from the linear ephemeris) times of transit for \Kb . In red, the measurements from \cite{Ballard2011}, in black our new measurements for all the {\it Kepler} quarters. In the lower plot the difference between the two measurements is shown for the data points in common, with the error bars obtained by summing in quadrature the errors from the two estimates. The small scatter of the residuals with respect to the size of the error bars demonstrate that we are not influenced by the exact methodology used to measure the $T_0$s.}
\label{fig:T0_difference_plot}
\end{figure}

Transit times from LC and SC light curves were matched together, keeping the SC measurements when available.  Transit times are reported in Table \ref{table:T0_table}. A comparison with B11 measurements of the observed minus predicted time of transit ($O-C$), using their linear ephemeris for both datasets, is shown in Figure \ref{fig:T0_difference_plot}. The scatter of the residuals, well within the error bars, shows that the methodologies are perfectly consistent, \ie , that we are limited by photon noise, data sampling and/or unknown systematics rather than the exact procedure followed to measure the transit times.

Due to an error in {\it Kepler} archiving system, at the time of B11 publication the time stamps of all the {\it Kepler} light curves were reported in Coordinated Universal Time system (UTC) instead of the Barycentric Dynamical Time system (TDB)\footnote{\url{http://archive.stsci.edu/kepler/timing_error.html}}. While this error was not affecting the internal consistency of B11 analysis, it must be taken into account when comparing time-series with timing accuracy better than a few minutes.  We corrected for this error before comparing B11 data with our new $T_0$ measurements.

\begin{table}
  \caption{Transit times of \Ks\ from Q0-Q17. }
\label{table:T0_table}      
\centering                                      
\begin{tabular}{l c c}          
\tableline\tableline                        

  Transit Number  & T$_0$ [BJD$_{\rm UTC}$] & $\sigma _{\rm T_0}$ [d]  \\

\tableline                                 

   0 &    2454959.7074 &   0.0014 \\
   1 &    2454968.9935 &   0.0023 \\
   2 &    2454978.2801 &   0.0020 \\
  ... &    ... &   ... \\

\tableline                        
\tableline
\end{tabular}
  \tablecomments{Table \ref{table:T0_table} is published in its entirety in the machine-readable format. A portion is shown here for guidance regarding its form and content.}
\end{table}

\section{Physical parameters}\label{sec:physical-parameters}

Atmospheric stellar parameters of \Ks\ were determined in B11 using Spectroscopy Made Easy ({\tt SME}, \citealt{Valenti2005}). Since this method may suffer of correlation between derived parameters \citep{Torres2012}, and having at our disposal several high-resolution spectra from HARPS-N, we decided to carry out an independent determination with an alternative approach, \ie , equivalent width measurements of individual spectral lines instead of fitting of the whole spectrum. We used all the spectra free from sky contamination to obtain a coadded spectrum with an average \snr\ of 350.

Stellar atmospheric parameters were determined using the classical line-of-growth approach. For this purpose we used the 2014 version of the line analysis and synthetic code {\tt MOOG}\footnote{Available at   \url{http://www.as.utexas.edu/~chris/moog.html}} \citep{Sneden1973}, which works in the assumption of local thermodynamic equilibrium (LTE), and the {\tt ATLAS9} grid of stellar model atmosphere from \cite{Castelli2004} with the new opacity distribution functions and no convective overshooting. Equivalent Width measurements were carried out with the code {\tt  ARESv2}\footnote{Available at \url{http://www.astro.up.pt/~sousasag/ares/}} \citep{Sousa2015} coupled with the updated linelist of \cite{Malavolta2016}, where the oscillator strength of the atomic lines have been modified to correctly take into account the chemical abundances from \cite{Asplund2009}.

Temperature and microturbulent velocity were determined by minimizing the trend of iron abundances from individual lines with respect to excitation potential and reduced equivalent width respectively, while the gravity $\log g$ was adjusted by imposing the same average abundance from neutral and ionized iron lines. For a detailed description of the procedure for the atmospheric parameters and associated errors we refer the reader to \cite{Dumusque2014}.
The derived atmospheric parameters are summarized in Table \ref{table:starpams_table}.

Our stellar atmospheric parameters agree within the uncertainties with the ones determined by B11, including the surface gravity which is usually the parameter most difficult to derive, and there is only a difference of 3 K in \teff\ despite the use of two complementary approaches and independent datasets. For this reason we adopted their determination for mass and radius of the star and physical radius of \Kb\ based on light curve analysis.

\begin{table}
\caption{Astrophysical parameters of the star.}              
\label{table:starpams_table}      
\centering                                      
\begin{tabular}{l c c }          
\tableline\tableline                        

Parameter & B11 & This work \\

\tableline                                 

\teff\ [K]            &     $5541 \pm 60$       & $5544 \pm 20$    \\
$\log g$               &    $4.59 \pm 0.10$ &     $4.51 \pm 0.03$    \\
\vmicro\  [\kms ]  & -  & $0.88 \pm 0.05$    \\
\gfeh               &   $-0.13 \pm 0.06 $   & $-0.08 \pm 0.02$  \\
\mstar\ [\msun ]   & $0.936 \pm 0.040$ & -  \\
\rstar\ [\rsun ]   & $0.859 \pm 0.018$ & - \\
  Age (Gyr)        & $1.9 \pm 1.7$ & -  \\
  \logRHK  & $-4.95 \pm 0.05 $ &  $-5.00 \pm 0.04 $  \\

\tableline                        
\tableline
\end{tabular}
\end{table}

\section{Stellar activity}\label{sec:stellar_activity}
\begin{figure}[htbp]
\includegraphics[width=\linewidth]{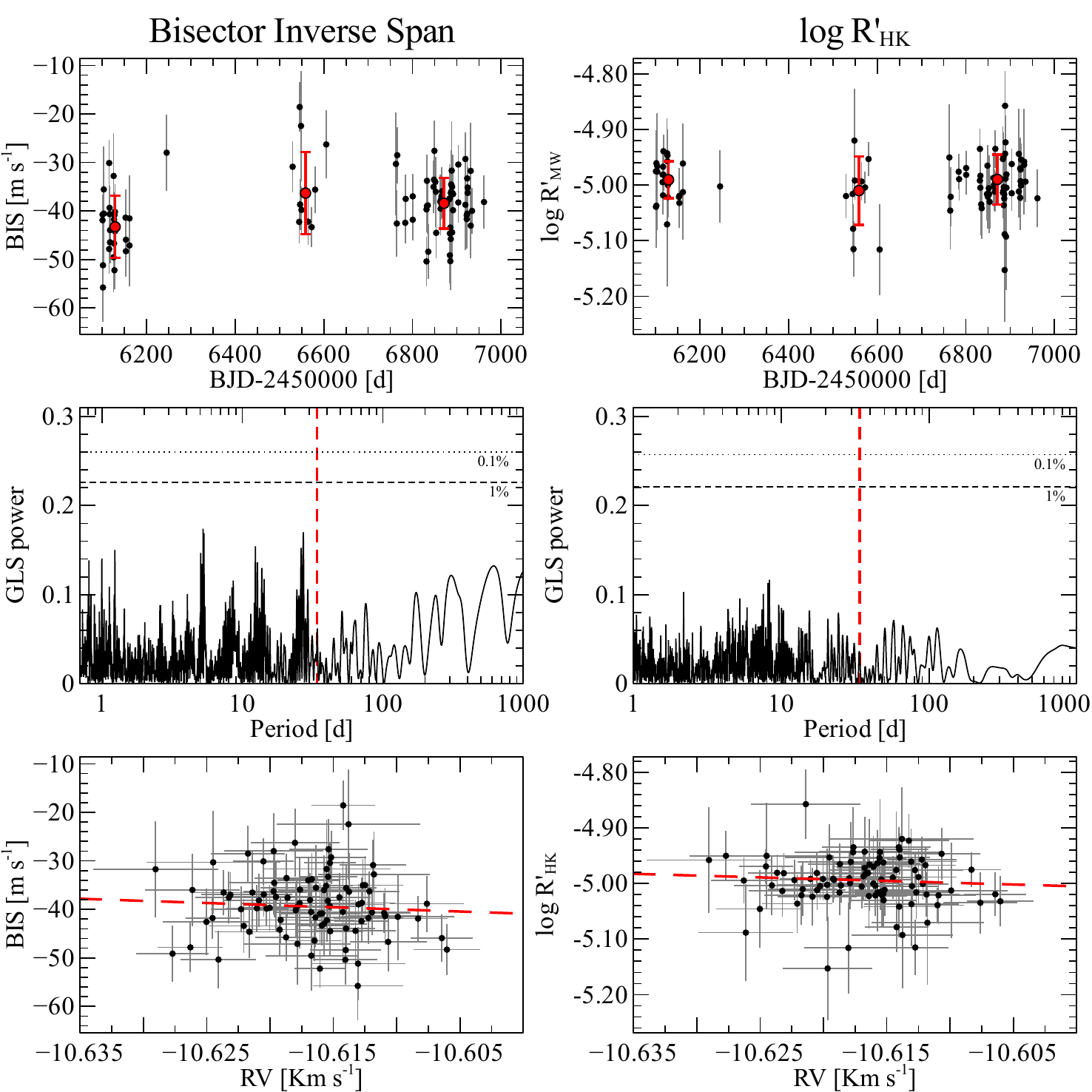}
\caption{The bisector inverse span (panels on the left side) and the \logRHK\ index (panels on the right side) are shown as example of the analysis conducted on CCF asymmetry and activity indexes. Upper panels: indexes as a function of time, the seasonal medians with the first and third quartiles indicated in red. Middle panels: GLS periodograms of the indexes, the rotational period of the star is indicated with a red vertical line. The 1\% and 0.1\% FAP levels are displayed as dashed and dotted horizontal lines, respectively. Lower panels: indicators as a function of RV. The best fit is represented by the dashed red line.}
\label{fig:activity_plot}
\end{figure}

Recently a considerable effort has been devoted to analyze the effect of stellar activity on RV measurements as well as $T_0$ determination \citep{Mazeh2015,Ioannidis2016}. The HARPS-N DRS automatically delivers several diagnostics for activity such as the Full Width Half Maximum of the CCF, the bisector inverse span and the \logRHK\ index, while several other chromospheric indexes such as the  H$\alpha$ index \citep{GomesDaSilva2011,Robertson2013} can be determined from the spectra themselves\footnote{The code to retrieve the activity indexes is available at \url{https://github.com/LucaMalavolta/}}. In Figure \ref{fig:activity_plot} the analysis of  BIS and \logRHK\ are reported as representative of all the indexes. For each index we checked the presence of any correlation with time, either by visual inspection (upper panels of Figure \ref{fig:activity_plot}) and with the  Generalized Lomb-Scargle (GLS) periodogram \citep{Zechmeister2009}, where the 1\% and 0.1\% false alarm probability (FAP) have been computed with a bootstrap approach (middle panels of Figure \ref{fig:activity_plot}). Finally, the presence of any correlation with RVs was verified by calculating the Spearman’s rank correlation coefficient $\rho $, the slope of the linear fit $m$ with its error and the p-value using the weighted least-square regression\footnote{StatsModels available at \url{http://statsmodels.sourceforge.net/}} (lower panels of Figure \ref{fig:activity_plot}). We omitted the FWHM from the analysis since a few changes of the spectrograph focus during the first year of observations modified the instrumental profile, hence the measured FWHM, without however affecting the measured RVs.

The absence of significant peaks in the periodogram of the indexes under analysis around the expected rotational period of 32 days (from B11 following \citealt{Noyes1984}), $34 \pm 6$ days and $36 \pm 3$ days (from HARPS-N \logRHK , following \citealt{Noyes1984} and \citealt{Mamajek2008} respectively), and the lack of statistically significant correlations between the indexes and RVs, confirmed the low activity level  of the star already deduced by B11 and consistent with \logRHK\ $ = -5.00 \pm 0.04 $\footnote{Note however that this value could be affected by interstellar medium absorption and be higher than measured \citep{Fossati2017}.}.

We also searched for stellar variability on the most recent Kepler Pre-search Data Conditioning (PDC) light curve, which presents several improvements in correcting instrumental trends (and thus is better suited to search for activity modulation) with respect to the light curve available at the time of B11 publication. For most of the time the star is photometrically quiet, while in some parts of the light curve a clear signal, likely due to stellar activity, is detected.  We applied the autocorrelation function technique over these portions of light curve and  we estimated a rotational period of 30 days. This signal is characterized by a short time scale of decay (a few rotational periods) and a rapid loss in coherence. We comment on the impact on RV in section \ref{sec:rv_analysis}.
  
\section{TTV  analysis}\label{sec:rv_ttv_analysis}

Dynamical analysis of the system was performed with {\tt  TRADES}\footnote{{\tt TRADES} is available at \url{https://github.com/lucaborsato/trades}} (TRAnsits and Dynamics of Exoplanetary Systems, \citealt{Borsato2014}), a N-body integrator with the capability of fitting RVs and $T_0$s simultaneously to determine the orbital parameters of the system through $\chi^2$ minimization.

Since only the innermost planet is transiting, it is extremely difficult to constrain the orbital parameters of all the planets in the system by TTV alone. In fact, attempts to fit the $T_0$s with a two-planet model resulted in a strong degeneracy between the mass and the period of the second planet, \ie , O-C diagrams with similar shape could be produced by jointly increasing the mass and the period of \Kc . The amplitude and shape of the $T_0$s however can still give us upper limits on the mass and period of the non-transiting planet if we compare the outcome of the dynamical simulations with the maximum mass compatible with the observed semi-amplitude of the RVs.

We proceeded as follows. We used {\tt TRADES} to perform a fit of the $T_0$s by assuming a two-planet model and fixing the mass of \Kb\ to a grid of values between 2.5 and 20 \mearth\ and a spacing of 2.5 \mearth , with its period already measured from the {\it Kepler} observations. To each point of this grid we assigned several values for the mass of \Kc\  randomly selected between 5 and 250 \mearth . For each combination of \Kb\ and \Kc\ masses, the other orbital parameters of the system were left free to vary and their best-fitting values were determined by $\chi^2$ minimization through the Levenberg-Marquardt algorithm. We discarded those solutions with at least one planet having eccentricity greater than 0.3, assuming that interacting planets meeting this condition  are likely to be unstable.

\begin{figure}[t]
\includegraphics[width=\linewidth]{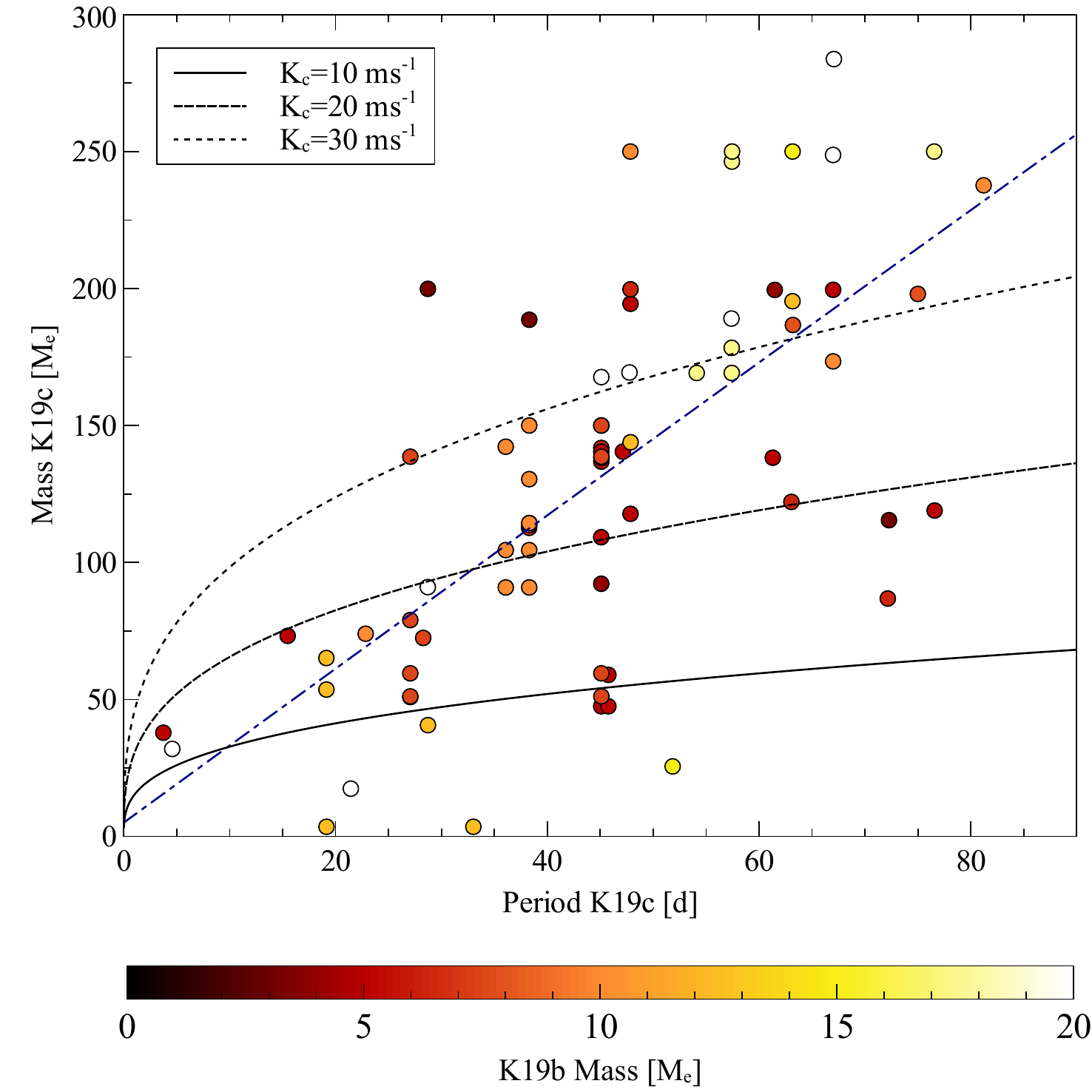}
\caption{The mass of \Kc\ versus its period, determined by fitting \Kb\ $T_0$s for a grid of masses of the two planets. The three lines represent the expected RV semi-amplitude of \Kc\  only as a function of its mass and period. A linear fit to the data is marked with a dash-dot blue line.}
\label{fig:ttb_lower_limits}
\end{figure}

In Figure \ref{fig:ttb_lower_limits} we show the results obtained for the mass of \Kc\ as a function of its period. The expected RV semi-amplitude of \Kc\ as a function of mass and period are superimposed. We note that many of the orbital configurations reported in the plot could be unstable, since dynamical stability was not checked yet at this stage (stability analysis is introduced in Section~\ref{sec:combined-rv-ttv}).
We can attempt to estimate a lower limit to the periods and masses of the non-transiting planets according to the observed semi-amplitude of the RVs, by taking advantage of the correlation between the mass and period of \Kc .
Our RVs have a peak-to-peak variation of 23 \ms , so if we keep into account the additional signal of \Kb\ (a few \ms\ in the case of a Neptune-like density), \Kc\ amplitude should necessarily lie below the $K=10$ \ms\ line. 
This fact suggests that a short period ( $\lesssim 50$ days) for \Kc\ should be expected, while nothing can be said regarding \Kb\ from TTV alone. We remind the reader that this analysis cannot be considered conclusive due to the reduced number of points and their scatter around the best linear fit.

\section{RV analysis}\label{sec:rv_analysis}
\begin{figure}
\includegraphics[width=\linewidth]{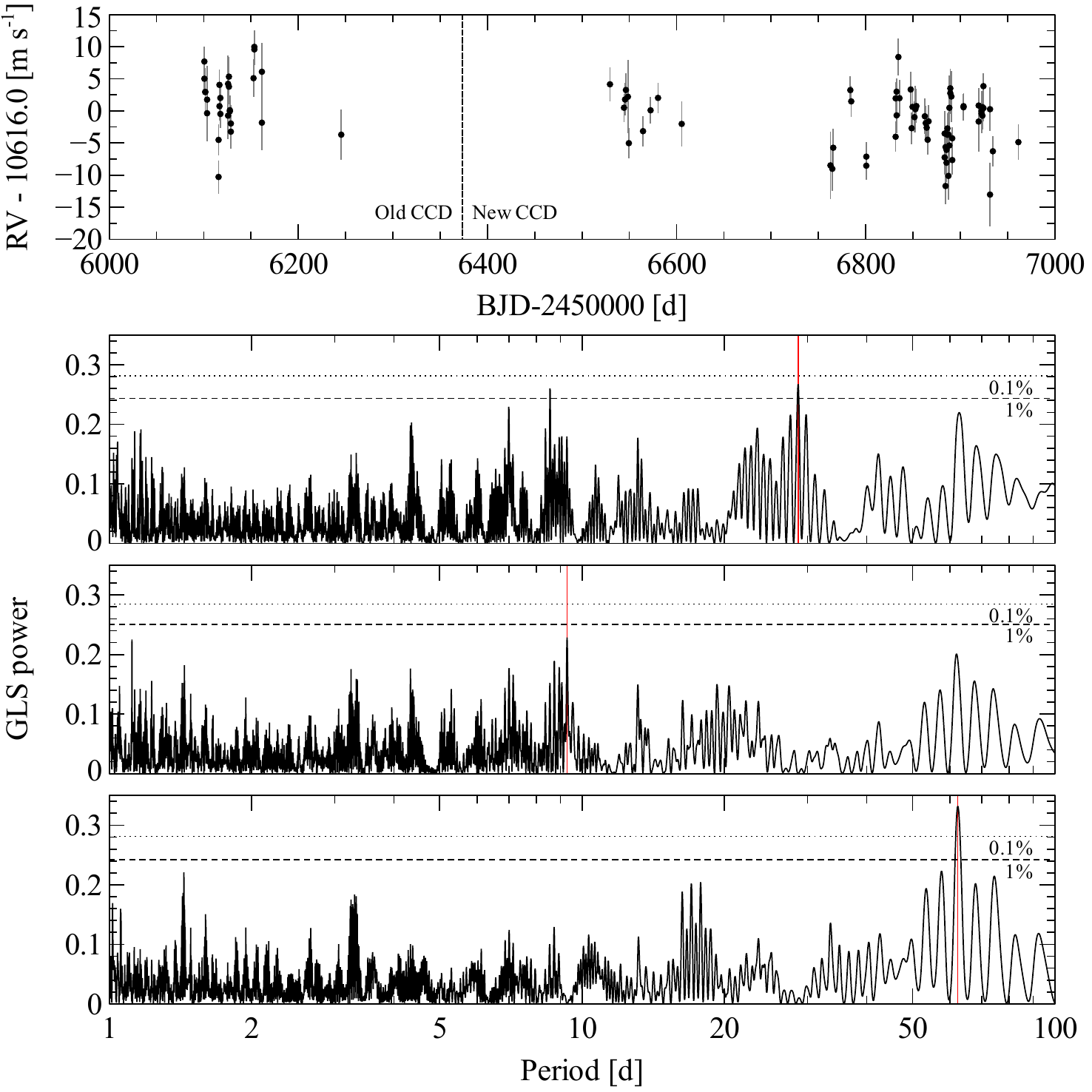}
\caption{Upper panel: the RV dataset. The vertical line divides data taken before and after the substitution of HARPS-N CCD. Lower panels: GLS periodograms computed starting from the initial dataset and iterating over the residuals. The 1\% and 0.1\% false alarm probabilities were estimated following a bootstrap approach for each periodogram. Red lines mark the best-fit period at 28.6 days, 9.29 days and 62.3 days.}
\label{fig:GLS_plot}
\end{figure}

While only one planet is transiting \Ks , the presence of at least one additional planet was inferred by the presence of TTVs. To reveal such planets, we first performed a frequentist analysis on the RV dataset by computing the GLS periodogram and iterating over the residuals until no significant periodicity was present. This analysis revealed two signals at 28.6 days and 62.3 days with false alarm probability lower than 1\% , while the signal at 9.3 days corresponding to \Kb\ was barely detected  (Figure \ref{fig:GLS_plot}).

The signal at 28 days is very close to the second order 3:1 mean motion resonance (MMR), which is among the possible configurations listed by B11 as the cause of the TTVs of \Kb. While the false alarm probability of this signal is below the traditional threshold for RV planet detection claims ($ \simeq 10^{-3}$), we can compare it with the probability of observing \Kc\ near another MMR resonance.  In order to do so, we scrambled once again the RV observations and calculated the fraction of periodograms which had a stronger peak with respect to the untouched dataset, in a frequency range of 5\% around the interior and exterior 1:2, 2:3, 3:4 first order resonances, the 1:3 and 3:5 second order resonances, and the 1:4 and 1:5 higher order resonances. The FAP of the signal at 28 days computed in this way decreases to $8 \times 10^{-4}$, \ie , it is unlikely that the signal at 28 days is a spurious signal caused by a planet in another MMR configuration.

The signal at $\simeq 28.6 $ days is consistent with the  rotational period of the star obtained from the active regions on {\it Kepler} light curve, which however have the characteristic of rapidly decaying and reappearing later at different phase and intensity. We checked if the RV signal has the same properties by performing a jacknife analysis by splitting the RV dataset in two parts (at BJD $= 2456700 $d) and determining the phase and amplitude of the signal on each dataset\footnote{We imposed as a prior the period P$=28.57 \pm 0.02$ day to compensate for the scarcity and poor sampling of the split datasets}. We obtained $ \mathrm{K} = 2.6 \pm 1.1 $ \ms,  $\phi = 3.0 \pm 0.3$ rad for the first half of the dataset, and  $ \mathrm{K} = 3.0 \pm 0.6 $ \ms,  $\phi = 2.8 \pm 0.2$ rad  for the second half, \ie , the signal is stable over several years, and therefore unlikely to be due to stellar activity. As an additional check, we computed the stacked Bayesian Generalized Lomb-Scargle (BGLS) periodogram \citep{Mortier2017}. The \snr\ is increasing with the square of the number of observations, as expected for a coherent (planetary) signal, although the plot is heavily affected by the poor sampling and the signal of \Kb\ near the 3:1 MMR, as we verified by performing the same analysis on synthetic datasets with the same temporal sampling. The absence of RV modulation due to stellar activity is supported by the analysis performed in Section~\ref{sec:stellar_activity}, where no periodicity or correlation with RVs is seen for any of the activity indexes.

The semi-amplitudes of these signals however cannot be inferred by the frequentist analysis alone, since an offset in RV between the data taken before and after September 2012 may exist due to the failure of the first CCD of HARPS-N (see \citealt{Bonomo2014}). The value of this offset cannot be determined a priori, even when observations of RV standard stars are available, since it may  depends primarily on the spectral type of the star (\ie , the two CCD may have a different efficiency as a function of wavelength). Introducing an RV offset as a free parameter is a common procedure for non-overlapping datasets, \eg , see \cite{Benatti2016}.

We then performed a tentative fit using the Markov-Chain Monte Carlo (MCMC) code {\tt PyORBIT}\footnote{Available at \url{https://github.com/LucaMalavolta/PyORBIT}} \citep{Malavolta2016}, allowing for two different systemic velocity $\gamma$ of the star and using a 3-planet model to fit the data.
We attempted two different fits with both circular and Keplerian orbits. For planet $c$ and $d$ we set
uniform priors in the range $[10, 50]$ days and $[50, 90]$ days respectively, while the other parameters were left to vary within their physically meaningful range (\eg , positive-definite RV semi-amplitude). The results are shown in Table \ref{table:pams_table_RV}.

Eccentricity and argument of pericenter of each of the three planets are poorly constrained by the RVs alone, thus affecting the precision on mass of the presumed planets. A combined analysis of RVs and TTV is then required to unambiguously detect and characterize the planets in the system.

\begin{table}
\caption{Orbital parameters for a 3-planet model for the \Ks\ system obatined from RV only, except for the period of \Kc\ which is constrained by the {\it Kepler} light curve. The reference time for the orbital elements is  T$_{\rm ref} = 2456624.82263024$ days.}
\label{table:pams_table_RV}      
\centering                                      
\begin{tabular}{l c c c l }          
\tableline\tableline                        

 & \Kb & \Kc & \Kd  \\     

\tableline                                 
 & \multicolumn{3}{c}{Circular orbits} \\
\tableline

             Period [d] & $   9.28699 \pm 10^{-5}  $ &  $      28.61    \pm   0.24     $  &  $    63.0      \pm  0.3     $ \\
          K [\ms ] & $ 2.3 \pm 0.5            $ & $  1.7 \pm 0.8 $                   &   $ 3.8 \pm 0.6 $ \\
$\phi $ [deg]           & $ 194.4 \pm 0.3 $ &  $ 185 \pm 43 $ & $174 \pm 6 $ \\
        Mass [\mearth ] & $   7.4   \pm  1.7    $ &  $      7.8     \pm   3.6      $  &  $    22.5      \pm  3.8     $ \\

\tableline                                 
 & \multicolumn{3}{c}{Keplerian orbits} \\
\tableline

             Period [d] & $   9.28699 \pm 10^{-5}  $ &  $ 28.54 \pm 0.27  $  &  $   62.9 \pm 0.3      $ \\
               K [\ms ] & $  2.6 \pm 0.6            $ &  $  1.95 \pm 0.8   $  &  $  4.0  \pm 0.7 $       \\
          $\phi $ [deg] & $  198 \pm  8             $ &  $  192  \pm   29  $  &  $    172   \pm 8      $ \\
$\sqrt{e} \cos \omega $ & $  0.0 \pm 0.2            $ &  $  0.0  \pm   0.4 $  &  $  0.31^{+0.20}_{-0.31} $ \\
$\sqrt{e} \sin \omega $ & $  0.0 \pm 0.3            $ &  $  0.0  \pm   0.3 $  &  $ -0.25^{0.35}_{-0.22}  $ \\
        Mass [\mearth ] & $   8.0   \pm  1.8        $ &  $ 8.6   \pm   3.5 $  &  $    23.1 \pm 3.8     $ \\
                    $e$ & $   \leq 0.13             $ &  $       \leq 0.30 $  &  $    \leq 0.32  $      \\
         $\omega$ [deg] & unconstrained               &  unconstrained        &  $ -36^{+65}_{-35} $      \\

\tableline

\end{tabular}
\end{table}

\section{Combined RV and TTV analysis}\label{sec:combined-rv-ttv}

The amplitude of dynamical perturbations between planets is very sensitive to the eccentricity and angular parameters (\ie , argument of pericenter $\omega$ and mean anomaly at reference time $\mathcal{M}_0$) of the planetary orbits, which however can be only poorly constrained  by the RVs, especially for planets in nearly circular orbits. For this reason we expect an overall improvement on the precision of the orbital parameters by simultaneously fitting RVs and TTVs.

Dynamical simulations are extremely time-consuming, and we have to use all the information at our disposal to reduce the extension of the parameter space. We used the results from the RV fit in Section~\ref{sec:rv_analysis} to put a constraint to the range of period, mass and orbital phase of each planet.

We followed an iterative approach to avoid being trapped in a local minimum of the $\chi ^2$. We started 10 separate runs of {\tt TRADES} with the Particle Swarm Optimization algorithm and loose priors on the periods ($\pm 10$ days around the expected period of each planet from the RV analysis), masses (between 0 and 40 \mearth ) and eccentricities ($e \leq 0.5$),
taking into account the limits imposed by photometry and RVs. We checked the stability of the outcome of each run and then we ran {\tt TRADES} again on a range of parameters which was half the size than in the previous run and centered of the outcome of the previous run with the lowest $\chi^2$ among those that satisfied the stability requirement. Convergence was considered achieved when all the runs resulted in similar parameters (within 5\% from the mean) and similar $\chi^2$ (10\% from the mean).

Following \cite{Gladman1993}, during the numerical integration we checked the stability criterion for each pair of planets: $\Delta = 2 \sqrt{3} R_H(i,j)$, where $\Delta = a_j-a_i$ is the semi-major axis difference between the j$_{\rm th}$ and  i$_{\rm th}$ planet, and $ R_H(i,j)$ is the mutual Hill radius between planet {\it i} and {\it j}.
At the end of each {\tt TRADES} fit we performed a N-body integration with {\tt  SymBA} \citep{Duncan1998} and  checked the stability of the result with the Frequency Map Analysis tool (FMA, \citealt{Laskar1992,Laskar1993a,Laskar1993b}) with the prescriptions of \cite{Marzari2002}. A system is considered stable if the coefficient of diffusion is lower than $10^{-5}$, in an unstable or chaotic state otherwise.

We used the value measured by B11 for the inclination $i$ of \Kb . Several attempts to fit the mutual inclinations of the non-transiting planets along with the other parameters resulted always in unstable solutions with high mutual inclinations. We took advantage of the low mutual inclinations of planetary orbital planes inferred from {\it Kepler} multi-planet systems \citep{Fabrycky2014} and the additional information coming from systems characterized with high-precision RVs \citep{Figueira2012} to impose coplanarity with \Kb\ for the other planets, while the longitude of the ascending node $\Omega$ was fixed to zero for all the planets. This assumption allowed us to drastically reduce computational time. The orbital period and mass of the planet, the eccentricity $e$, the argument of periastron $\omega$ and the mean anomaly at the reference epoch $\mathcal{M}$ were left as free parameters.

Differently from \citep{Borsato2014}, we fitted  $\sqrt{e} \cos{\omega} $ and $\sqrt{e} \sin{\omega}$ instead of $e \cos{\omega} $ and $e \sin{\omega}$, and the mean longitude at the reference epoch $\lambda = \omega +  \mathcal{M} + \Omega$ (where $\Omega$ is the longitude of the ascending node, fixed to zero for being unconstrained by the data) instead of $\mathcal{M}$. We used scaled stellar and planetary masses as commonly done in TTV analysis (\eg , \citealt{Nesvorny2012}). A RV offset between the data taken before and after September 2012 was included as a free parameter to take into account the change of the CCD (see Section~\ref{sec:rv_analysis}). 

We used the solution obtained with the global exploration of the parameter space as a starting point for the Bayesian analysis. For this purpose we expanded {\tt TRADES} functionalities with the {\tt emcee} package \citep{ForemanMackey2013}, an affine invariant Markov chain Monte Carlo (MCMC) ensemble sampler, and made it available to the community in the {\tt TRADES} repository. Following \cite{Feigelson2012}, we calculated the log-likelihood $\ln \mathcal{L}$ from the $\chi^2$ using Equation \ref{eq:chi2_to_loglike}, where $dof$ is the degree of the freedom of the problem.
\begin{equation}\label{eq:chi2_to_loglike}
  \ln \mathcal{L} =  -\ln{(2 \pi)}\frac{\textrm{dof}}{2} - \frac{\sum{\ln{\sigma^2}}}{2} - \frac{\chi^2}{2}
\end{equation}
  
We tested three different scenarios. The first model assumed that the signal at  $\simeq 28$ days (the first one being detected in the RV periodogram) is the only planet in the system other than  \Kb . The second model is still a two-planet model, but here we assumed that the signal at $\simeq 28$ days is due to activity (without any effects on the T$_{0}$s) and that the system consists of two planets at $\simeq 9.2$ and $\simeq 62$ days. The third model assumed that all the RV signals have planetary origin. The results are presented in the following subsections. We note that a planet in a strong MMR with \Kb\ could still produce the TTVs while having a RV semi-amplitude below our detection sensitivity. Following B11, the TTVs of \Kb\ could be explained by a planet in [2:1, 3:2, 4:3] MMRs with masses [4, 2, 1 ] \mearth\ respectively. The signal of the perturbing planet would nave an RV semi-amplitude on the order of [1, 0.6, 0.3] \ms , \ie , beyond the reach of modern velocimeters given the magnitude of our target. This scenario however requires an activity origin for the 28 days signals, which is not supported by the analysis of Section~\ref{sec:stellar_activity} and the coherent nature over years of the RV signal in opposition to the short time-scale decay of the spots observed in {\it Kepler} photometry, as shown in Section \ref{sec:rv_analysis}. For these reasons we can regard this scenario as unlikely.

\subsection{Two-planet model}\label{sec:two_planets_no_external_fit}
We tested the two-planet model following the steps described in Section~ \ref{sec:combined-rv-ttv}, with the only additional constraint of \Kc\ having a period lower than $40$ days. We ran the MCMC sampler for 50000 steps, with a number of chains twice the dimensionality of the problem. We checked the convergence of the chains using the Gelman-Rubin statistic ($\hat R < 1.03$, \citealt{Gelman1992,Ford2006}), and we built the posterior distributions with the last 20000 steps and a thinning factor of 200.

Results are listed in Table~\ref{table:pams_table_TTV}. Rather than using the median or the mode, we summarize the outcome of the analysis by selecting from the chains a sample with the nearest $\chi ^2$ to the median of its distribution and with each parameter within 
its confidence interval. Confidence intervals are computed by taking the 15.87th and 84.14th percentiles of the distributions, and are reported in the table with respect to the selected sample.

\subsection{Two-planet and  stellar activity model}\label{sec:two_planets_activity_fit}

The presence of the activity signal can potentially affect the analysis, but {\tt TRADES} is not equipped to deal with non-planetary signals in RV datasets. For this reason we decided to remove such signal from the RV time series before starting the global exploration of the parameter space, using as dataset the residuals of the first iteration of the GLS analysis in Section~\ref{sec:rv_analysis}. The global solution was used as starting point for a MCMC analysis with {\tt PyORBIT}, this time using the original RV dataset and the activity signal modeled with a Keplerian curve (as similarly done in \citealt{Pepe2013}). RVs and T$_{0}$s calculations regarding the two interacting planets were performed by calling the dynamical integrator of TRADES through a {\tt FORTRAN90} wrapper.
As additional test-cases, we run TRADES using the two-planet model and no activity modeling on both the original RV dataset {(imposing a lower limit of 40 days for the period of \Kc , to avoid falling in the case analyzed in Section~\ref{sec:two_planets_no_external_fit})} and the {\it GLS-cleaned} dataset. 
In all cases we repeated the analysis without imposing any constraint on eccentricity, since the eccentricities of both planets were moving towards the upper boundary imposed in Section~\ref{sec:combined-rv-ttv}. We followed the same methodology described in Section~\ref{sec:two_planets_no_external_fit} to run the MCMC and extract the results reported in Table.

The eccentricity of \Kb\, is extremely high in all the cases we considered. The most likely explanation is that high eccentricities for both planets are required to produce the same T$_{0}$s while keeping their masses within the boundaries set by the RVs.  Following \cite{Burke2008} we computed the ratio $\tau$ of the transit duration of an eccentric orbit with respect to a circular orbit, and we obtained  $\tau=0.64 \pm 0.04$ for the original RV dataset, $\tau=0.58 \pm 0.06$ for the {\it GLS-cleaned} case, and $0.54 \pm 0.10$ for the planets$+$activity case. All these values are well below the value of $\tau=0.7$ that B11 considered as the minimum reasonable value for the transit duration ratio due to the eccentricity of the planet from the analysis of the {\it Kepler} light curve. We note that the period-mass combinations we obtain for the outer planet fall in an empty region of Figure~\ref{fig:ttb_lower_limits}, which however was obtained by selecting those solutions with $e<0.3$ for at least one of the planets.

\subsection{Three-planet model}\label{sec:three_planets_fit}
 
Bayesian analysis for the three-planet model were performed using {\tt TRADES} combined with {\tt emcee} (see Section~\ref{sec:combined-rv-ttv}). From a preliminary analysis we noticed that the chains were affected by poor mixing, with a Gelman-Rubin statistic $\hat R \simeq 1.3$ \citep{Gelman1992,Ford2006}. We decided to run 100 chains for an extensive number of steps $(250000)$ to overcome the poor mixing and perform a proper exploration of the parameter space, knowing that we already reside near the global minimum of the $\chi^2$. After the first 150000 iterations we did not see any variation of the posterior distributions while increasing the length of the chains, making us confident of the robustness of our result. We finally built the posterior probability by drawing 40000 independent samples from the chains, after removing the burn-in part and applying a thinning factor equal to their auto-correlation time ($\simeq 100$ steps).

\begin{figure}
\includegraphics[width=\linewidth]{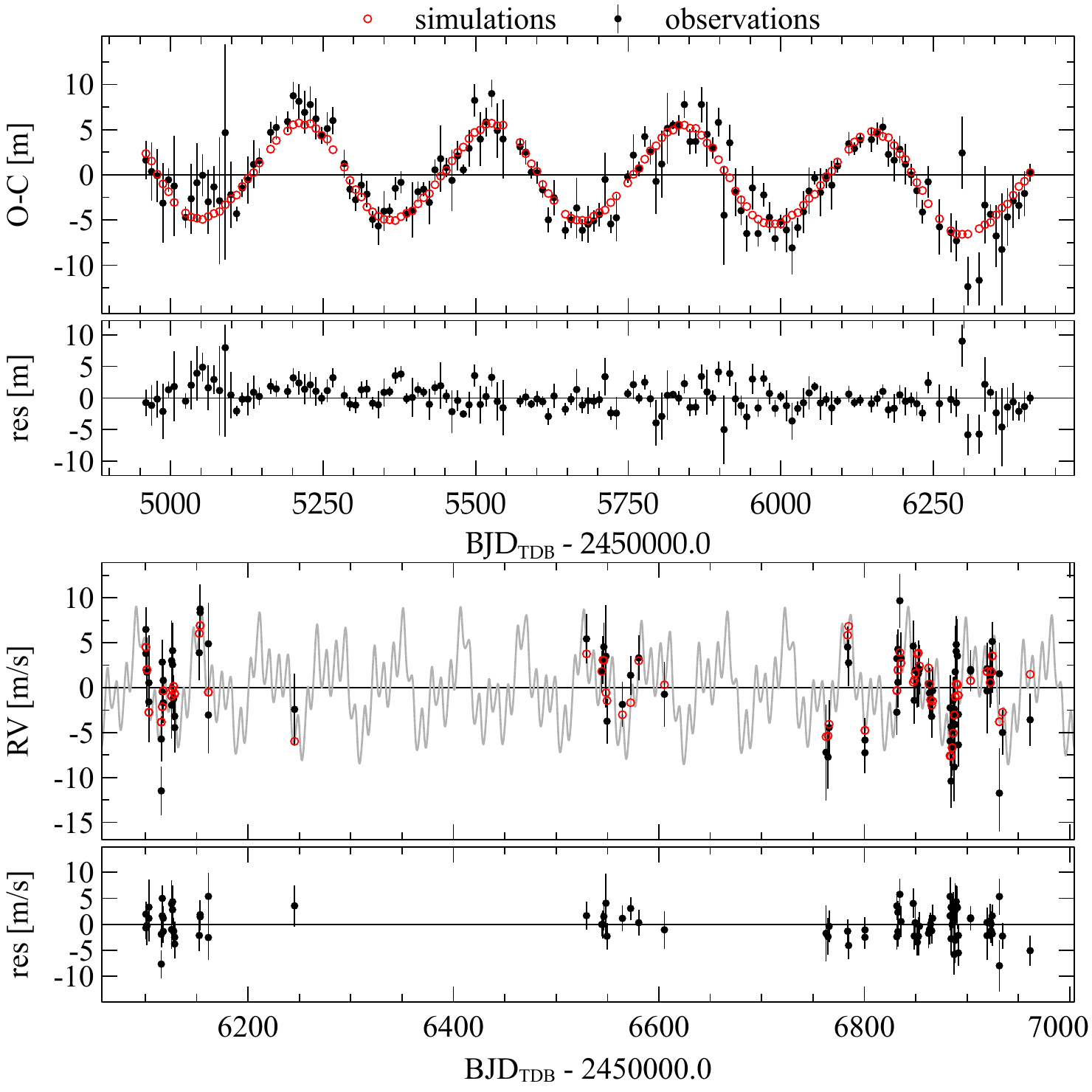}
\caption{Top panels: Observed-Calculated (O-C) transit times and residuals for \Kb . Lower panels: RVs and residuals of the \Ks system. Red empty circles represent the predicted values from the solution obtained with {\tt TRADES}. The radial velocity curve of the \Ks\ system obtained from the dynamical integration is shown in gray.}
\label{fig:TTV_RV_fit}
\end{figure}

Our results are listed in Table \ref{table:pams_table_TTV}. The confidence intervals of the posteriors are computed by taking the 15.87th and 84.14th percentiles of the distributions, and they are reported as error bars around a sample solution selected as in \ref{sec:two_planets_activity_fit}. From now on, we will use the reported values as a representative \textit{solution} of the orbital parameters of the planets. In Figure \ref{fig:TTV_RV_fit} we show the solution overplotted on $T_0$s and RVs, with their respective residuals.

\begin{figure*}
\plottwo{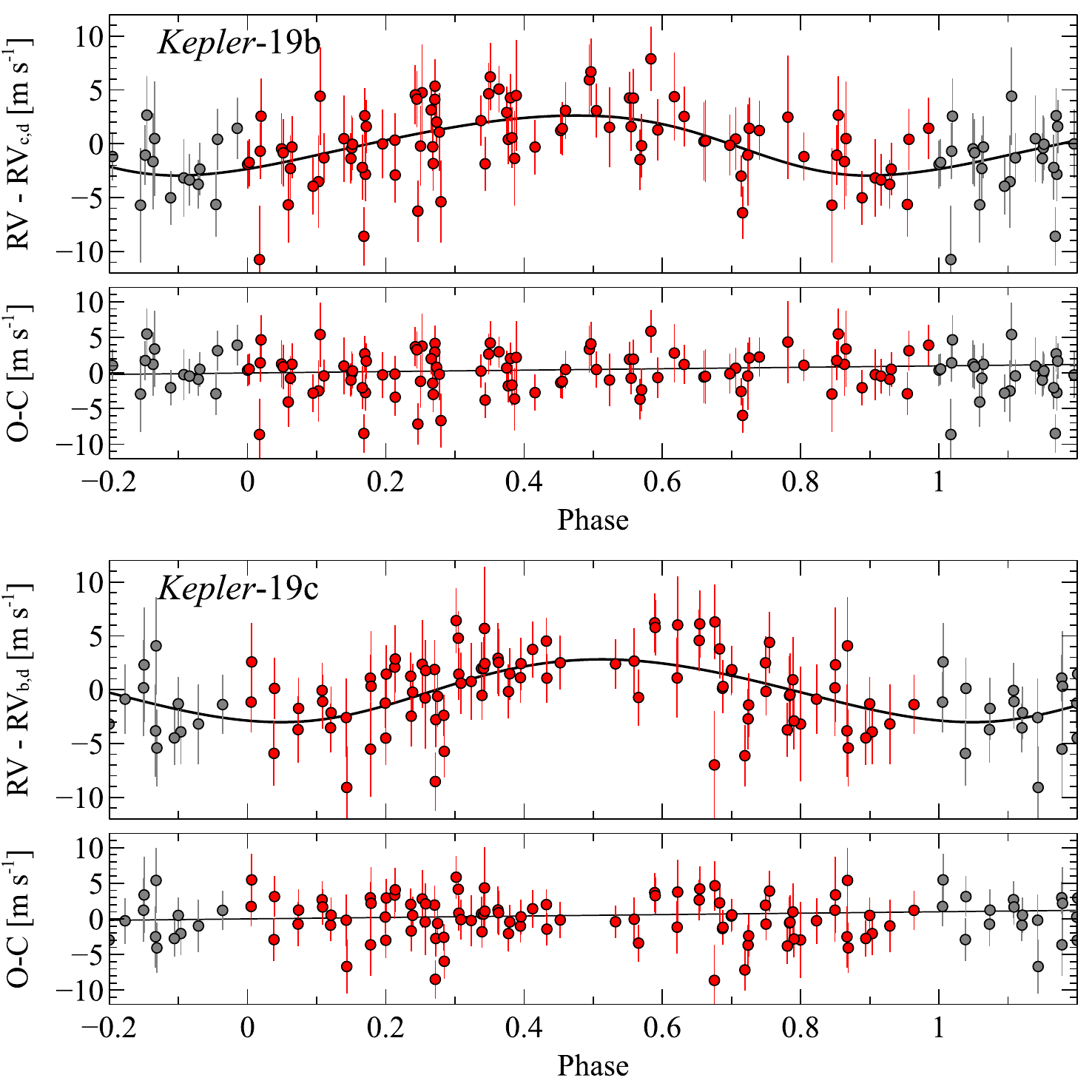}{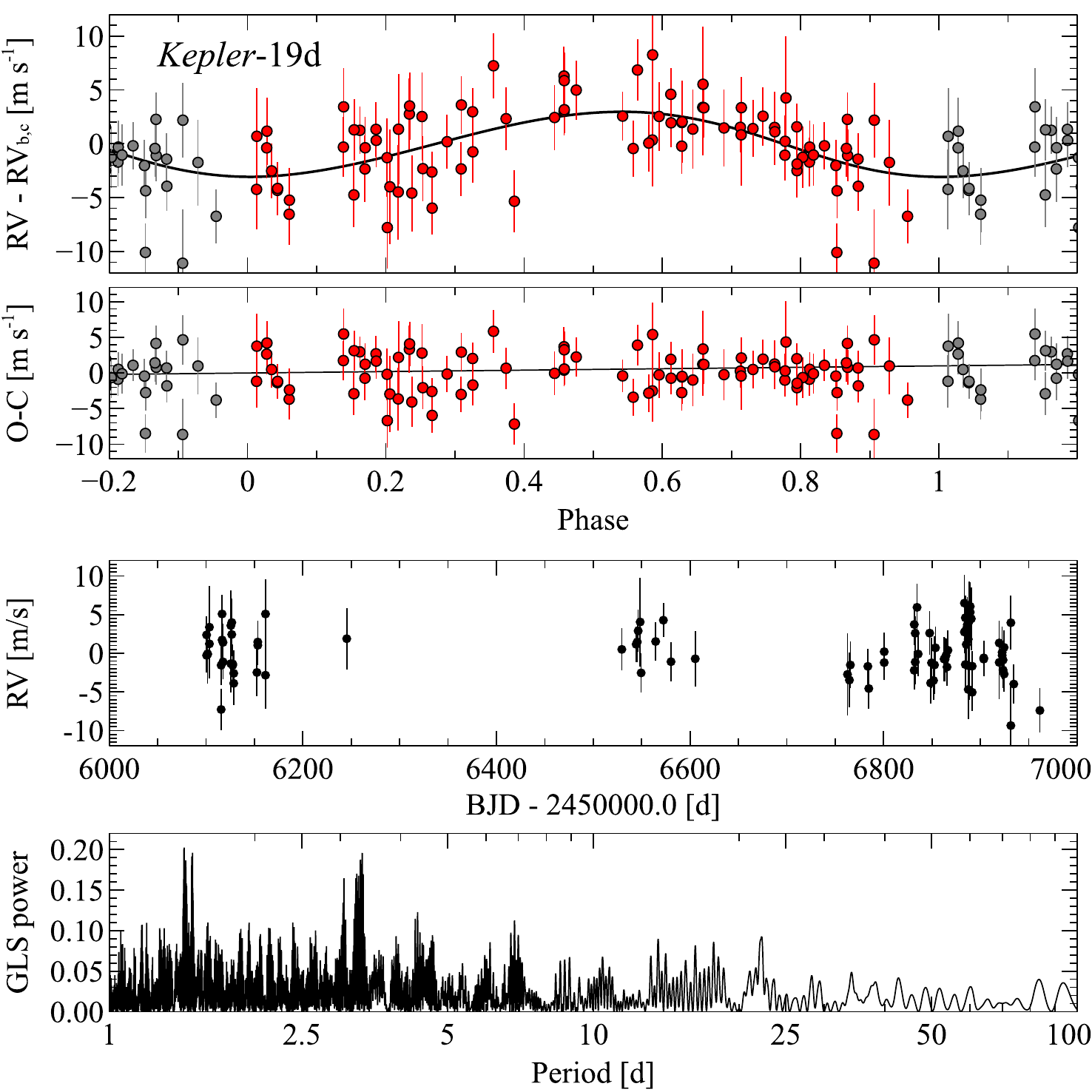}
\caption{Orbital solution and RV residuals for \Kb\ (upper-left panels), \Kc\ (lower-left panels) and \Kd\ (upper-right panels), phased on the period of the corresponding planet after removing the RV contribution from the other planets. These plots have been obtained using non-interacting Keplerian orbits, and are intended for illustrative purpose only. In the lower-right panels, both the RV residuals after subtracting the dynamical solution from {\tt TRADES} and their periodogram show no evidence for additional signals that are statistically significant.}
\label{fig:RV_phased}
\end{figure*}

In Table~\ref{table:model_selection} we compare the outcome of the different models under examination. The three-planet solution is the favorite one according to the BIC ( having a $\Delta \textrm{BIC}>10$ with respect to all the other models under analysis, \citealt{kassr95}), and the considerations at the end of Section~\ref{sec:two_planets_activity_fit} further strengthen this result. Our three-planet solution has a total reduced $\chi ^2_r = 1.25$ ($\chi^2 = 272.7$, $\textrm{BIC}=365.5$), where the contribution from RV and TTV is respectively 0.37 and 0.88\footnote{These two values do not correspond to the individual $\chi^2_{red}$ of RVs and TTVs, since the sum of residuals of each dataset is divided by the total number of data points.}. The standard deviation of the residual RVs is 2.9 \ms , consistent with the average error of the measurements, and while several peaks can be found in the periodogram, none of them reaches the 1\% false-alarm probability threshold (Figure \ref{fig:RV_phased}). Similarly, the standard deviation of the TTV of 2.2 minutes is consistent with the average error of 2.1 minutes.

For the three planets we determined their masses with a precision of $\sigma_{M_b} = 1.6 $ \mearth\ (error of 19\% on planetary mass) for \Kb ,  $\sigma_{M_c} = 2.7 $ \mearth\  (error of 21\%) for \Kc\ and $\sigma_{M_d} =  3.4$ \mearth\  (error of 17\%) for \Kd . We included in the computation the uncertainty on the stellar mass
and the effect of orbital inclinations, assuming as representative distributions for the latter a normal distribution with $i_c=i_d=89.94^\circ$ and dispersion $\sigma_i = 5 ^\circ$ for \Kc\ and $\sigma_i = 15.0 ^\circ$ for \Kd , after Section~\ref{sec:mutual-inclinations}. Knowing that \Kc\ and \Kd\ should have a transit depth greater than 0.5 mmag \citep{Winn2011_arXiv}, a visual inspection of the {\it Kepler} light-curve confirmed that the planets are not transiting.

Our solution agrees with the solution in Table \ref{table:pams_table_RV}, except for the mass of \Kc\ , with a 2-$\sigma$ difference between the RV-only and RV$+$TTV determinations. The origin of this discrepancy is likely due to the uncertainty in the orbital phase and eccentricity from the RV-only analysis, with one planet absorbing the signal of the other, possibly coupled with the effect described in Section~\ref{sec:dynamical-versus-non}.

The RV offset between the \textit{old} and \textit{new} CCD is $\simeq 2$ \ms, \ie , within the RV noise. 
We do not expect any influence of the RV offset on the outcome of the analysis, since the three planets have periods fully contained within the time span of both \textit{old} and \textit{new} CCD datasets.

\begin{table*}
  \caption{Orbital parameters of the planets in the \Ks\ system obatined from TTV+RV MCMC analysis using different assumptions for the number of planets and the stellar activity. The reference time for the orbital elements is  T$_{\rm ref} = 2456624.82263024$ d. }
\label{table:pams_table_TTV}      
\centering                                      
\begin{tabular}{l CCCCCCCC}          
\tableline                                 
\tableline                                 
 Planet &  \textrm{ Period [d]} & \textrm{Mass [\mearth ]} & \sqrt{e} \cos \omega  & \sqrt{e} \sin \omega  & \lambda \textrm{[deg]} & \textrm{e} &  \omega \textrm{[deg]} & \mathcal{M} \textrm{[deg]} \\
\tableline                                 
    \multicolumn{9}{c}{Two-planet model, P$_{c} < 40$ days} \\
\tableline
\Kb & 9.287108_{-0.00003}^{+0.00006} &  8.2_{-1.4}^{+1.6} &  0.23_{-0.02}^{+0.04} &  0.21_{-0.05}^{+0.03} &  188.9_{-1.5}^{+0.9} &  0.10_{-0.01}^{+0.01} &  42.5_{-11.6}^{+5.7} & 146.4_{-5.1}^{+10.5} \\
\Kc & 28.723_{-0.007}^{+0.0005} &  17.0_{-2.4}^{+2.4} &  0.48_{-0.04}^{+0.01} &  0.15_{-0.11}^{+0.02} &  175.5_{-7.0}^{+3.0} &  0.25_{-0.04}^{+0.01} & 17.3_{-12.7}^{+2.9} &  158.1_{-1.0}^{+6.8} \\
\tableline                                 
    \multicolumn{9}{c}{Two-planet model, P$_{c} > 40$ days} \\
\tableline
  \Kb &   9.286970_{-0.00005}^{+0.00009}   & 8.6 _{-1.7}^{+1.9} &   -0.09_{-0.06}^{+0.11}   &   0.64_{-0.05}^{+0.04}   &   201.6_{-6.3}^{+3.5  }   &   0.42 _{-0.05}^{+0.06}   &   97.9 _{-9.9}^{+5.5}   &   103.7 _{-2.2 }^{+3.6  }   \\
\Kc &   63.128 _{-0.007}^{+0.010}   & 21.7 _{-3.7}^{+1.8} &    -0.11 _{-0.02}^{+0.25}   &   -0.59 _{-0.02}^{+0.02}   &   167.7 _{-0.1  }^{+14.4 }   &   0.36_{-0.02  }^{+0.011}   &   -100.4_{-1.2}^{+23.9}   &   268.1_{-10.1}^{+0.2}   \\
\tableline
  \multicolumn{9}{c}{Two-planet model, {\it GLS-cleaned} RV dataset} \\
\tableline
\Kb &  9.28696_{-0.00006}^{+0.00006} & 8.4_{-1.6}^{+1.3} & -0.08_{-0.17}^{+0.02} & 0.74_{-0.11}^{+0.02} & 201.1_{-1.5}^{+10.2} & 0.551_{-0.11}^{+0.03} & 95.9_{-1.7}^{+14.8} & 105.2_{-4.8}^{+0.3} \\
\Kc &  63.128_{-0.008}^{+0.011} &  17.4_{-2.9}^{+2.9} &  -0.07_{-0.08}^{+0.11} & -0.57_{-0.03}^{+0.01} & 169.3_{-3.3}^{+10.2} & 0.332_{-0.006}^{+0.035} & -97.4_{-7.4}^{+10.9} & 266.7_{-3.4}^{+6.26} \\
\tableline
 \multicolumn{9}{c}{Two-planet and  activity model} \\
\tableline
\Kb &    9.28696_{-0.00004}^{+0.00008}  & 7.8_{-1.9}^{+2.1} &     -0.02_{-0.11}^{+0.19}  &    0.71_{-0.06}^{+0.09}  &    197.9_{-11.3}^{+6.4}  &    0.50_{-0.07}^{+0.17}  &    91.7_{-14.8}^{+9.2}  &    106.2_{-1.8}^{+2.6}  \\
Act &      28.57_{-0.11}^{+0.08}  & (2.3_{-0.3}^{+1.1})\tablenotemark{a} &   0.22_{-0.68}^{+0.31}  &    0.34_{-0.23}^{+0.31}  &    178.4_{-23.6}^{+2.0}  &    0.16_{0.01}^{+0.45}  &    56.4_{-36.1}^{+84.4}  &    122.0_{-86.7}^{+31.4}  \\
\Kc &     63.139_{-0.019}^{+0.003}  &  16.2_{-2.7}^{+2.4} &   0.11_{-0.24}^{+0.02}  &    -0.58_{-0.02}^{+0.04}  &    178.3_{-15.6}^{+1.16}  &    0.35_{-0.04}^{+0.02}  &    -79.0_{-23.6}^{+1.3}  &    257.4_{-1.7}^{+9.0}  \\
\tableline                                 
  \multicolumn{9}{c}{Three-planet model} \\
  \tableline
  \Kb & 9.28716_{-0.00006}^{+0.00004} & 8.4_{-1.5}^{+1.6} & 0.17_{-0.03}^{+0.05} & 0.29_{-0.06}^{+0.04} &  190.3_{-1.9}^{+1.0} &  0.12_{-0.02}^{+0.02} & 59.1_{-11.7}^{+68.9} & 131.2_{-6.1}^{+10.4} \\
\Kc & 28.731_{-0.005}^{+0.012} &  13.1_{-2.7}^{+2.7} & 0.42_{-0.03}^{+0.04} & 0.19_{-0.05}^{+0.10} & 181.5_{-4.5}^{+6.0} &  0.21_{-0.07}^{+0.05} & 23.8_{-6.8}^{+11.1} & 157.6_{-6.6}^{+3.1} \\
\Kd & 62.95_{-0.30}^{+0.04} & 22.5_{-5.6}^{+1.2} & 0.13_{-0.38}^{+0.17} & 0.18_{-0.19}^{+0.21} & 170.1_{-0.3}^{+12.3} & 0.05_{-0.01}^{+0.16} & 55.2_{-57.4}^{+76.8} & 114.9_{-70.5}^{+63.7} \\
\tableline
\end{tabular}
\tablenotetext{a}{Semi-amplitude of the signal in \ms}
\end{table*}

\begin{table*}
\caption{Statistical indexes for the different models under exams. The number of parameters in the fit, the degree of freedom ($dof$), the $\chi^2$ and its reduced value, the log-likelihood $\ln \mathcal{L}$ and the Bayesian Information Criterion (BIC) are reported. }
\label{table:model_selection}
\centering
\begin{tabular}{l CCCCCCC} 
\tableline
\tableline
Model & \textrm{N. parameters} & dof & \chi^2 & \chi^2_{\rm red} & \ln \mathcal{L} & \textrm{BIC} \\
  \tableline
  2 planets  P$_{c} < 40$ days& 12 & 223 &   323.8_{-7.4}^{+5.5} &  1.45_{-0.02}^{+0.01}&  498.6_{-3.7}^{+1.0}&  389.4_{-1.9}^{+5.5} \\
  2 planets, P$_{c} > 40$ days & 12 & 223 & 324.0_{-3.4}^{+4.4} & 1.45_{-0.01}^{+0.02} & 499.5_{-2.2}^{+1.6} & 389.5_{-3.4}^{+4.4} \\
  2 planets, {\it GLS-cleaned} RVs & 12 & 223 &  318.8_{-3.5}^{+4.4} & 1.43_{-0.02}^{+0.02} & 502.1_{-2.2}^{+1.8} & 384.3_{-3.5}^{+4.4} \\
  2 planets and activity & 17 & 218 & 319.5_{-6.1}^{+8.8} & 1.47_{-0.03}^{+0.04} & 506.3_{-4.4}^{+3.0} & 412.3_{-6.1}^{+8.8} \\
  3 planets   & 17 & 218 & 276.7_{-5.27}^{+6.9} & 1.27_{-0.02}^{+0.03} & 527.7_{-3.4}^{+2.6} & 369.5_{-5.3}^{+6.9} \\
\tableline
\end{tabular}
\end{table*}

\begin{figure*}
\includegraphics[width=\linewidth]{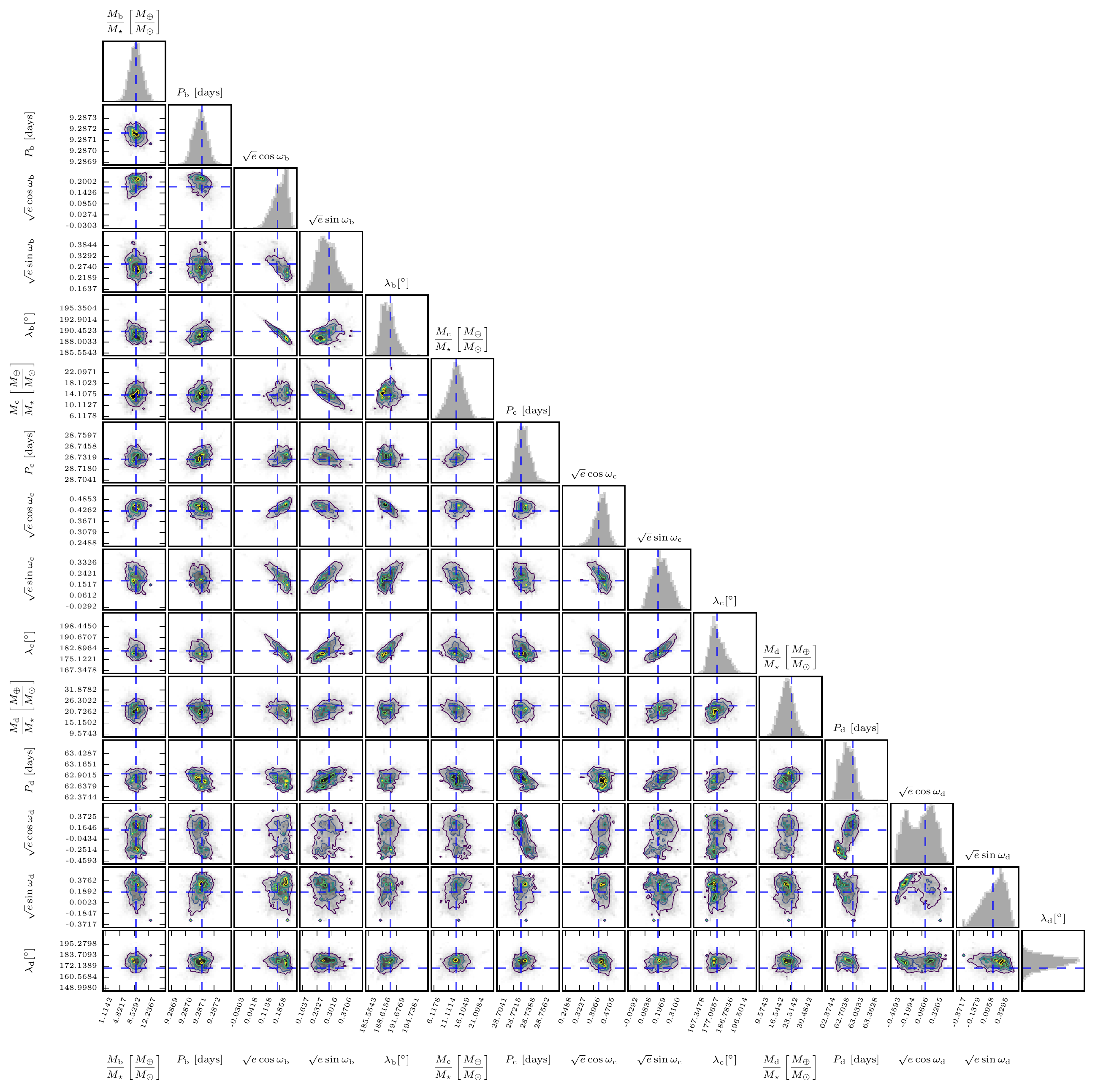}
\caption{ \bf Posterior distributions of the fitted parameters for the three-planet model for the \Ks\ system. Dashed blue lines identify the reference solution listed in  Table~\ref{table:pams_table_TTV}.}
\label{fig:corner_plot}
\end{figure*}

\section{Mutual inclinations}\label{sec:mutual-inclinations}
\begin{figure}[htbp]
\includegraphics[width=\linewidth]{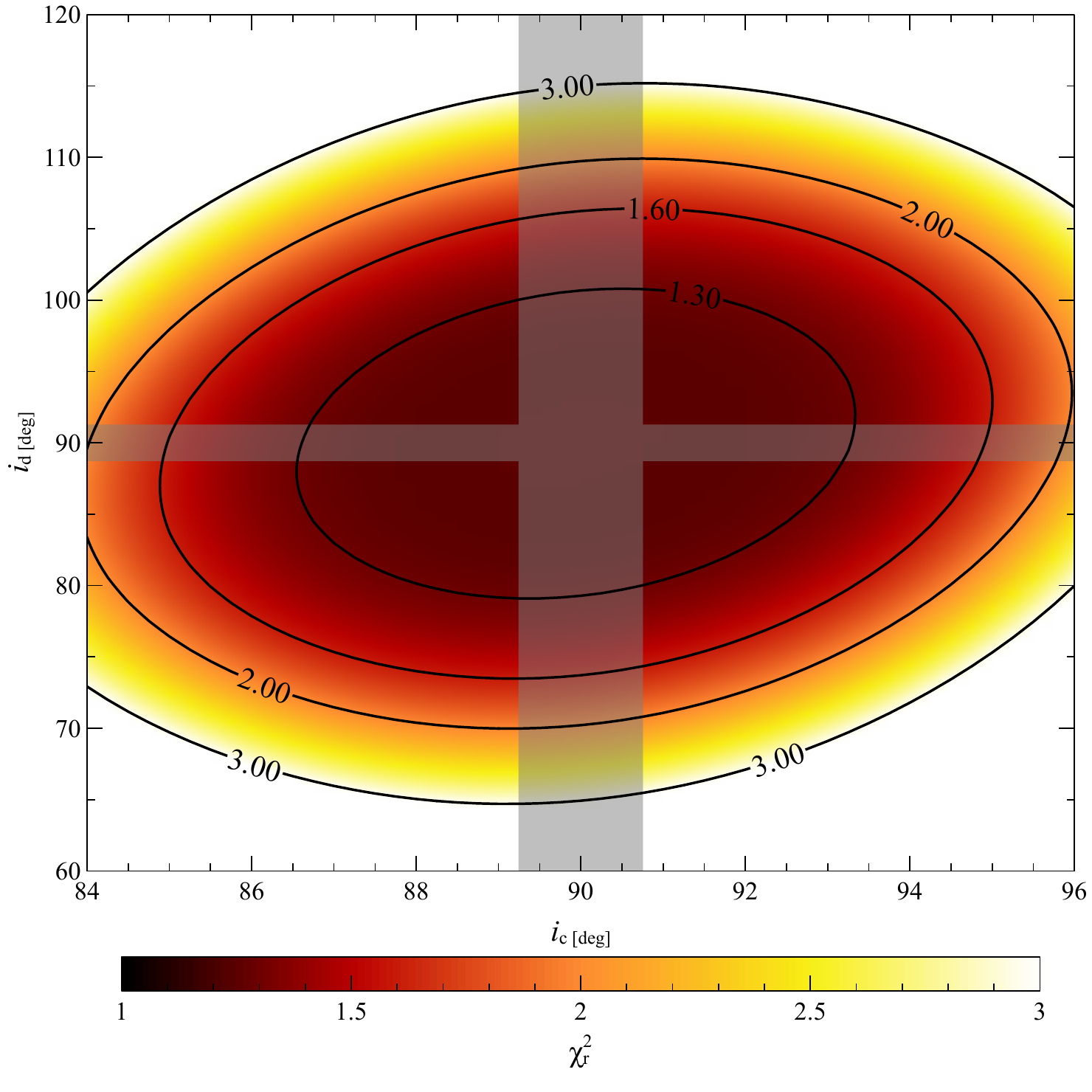}
\caption{Distribution of $\chi^2_{red}$ as a function of \Kc\ and \Kd\ inclinations, with the other orbital parameters fixed to our solution (Table \ref{table:pams_table_TTV}). Contour lines for several values of the $\chi^2_{red}$ are shown for reference. Values of inclination for which one of the planets would transit are shadowed in gray.}
\label{fig:K19cd_incl}
\end{figure}

In Section~\ref{sec:combined-rv-ttv} we assumed co-planarity with \Kb\ (the only planet with known inclination) to derive the orbital parameters of the additional planets in the system.  To check if this assumption was still valid after determining the orbital period of the additional planets, we determined the upper limit on inclination for which a transit is visible for a given planet, by inverting Equation 7 of \cite{Winn2011_arXiv} with the assumption for the impact parameter $b_{\rm tra}=1$. To properly take into account variation with time of $e$ and $\omega$ induced by dynamical interactions, we selected 1000 $(e , \omega )$ pairs for each planet by randomly sampling in time the integration our best solution over the {\it Kepler} observational time span. Error in the stellar radius was included by generating random samples from a normal distribution with mean \rstar\ and standard deviation $\sigma_{{\rm R}_{\star} }$ as reported in Table \ref{table:starpams_table}. We obtained $i_{\rm min}=88.72 \pm 0.03 ^\circ $ for \Kc\ and $i_{\rm min}=89.24 \pm 0.02 ^\circ $ for \Kd . Since we assumed  $i_b=89.94$ (from B11), then the coplanarity assumption cannot hold for \Kc .
We assessed the influence of orbital inclinations on the validity of our solution by running dynamical simulations for a grid of $i_c$ and $i_d$ (from 60 to 120 degrees, with step of $0.25^\circ$  for $i_c$ and $0.5^\circ$ for $i_d$), with the remaining orbital parameters fixed to our best solution, and determining the reduced $\chi^2$ as in Section~\ref{sec:combined-rv-ttv}. As shown in Figure \ref{fig:K19cd_incl}, the reduced $\chi^2$ increases rapidly with \Kc\  going farther from coplanarity with \Kb , reaching a value of $\approx 1.6$ for $|i_b-i_c| \simeq 5 ^\circ$, while \Kd\ can span a larger interval in inclinations without affecting the outcome (although increasing $|i_b-i_d|$ would affect negatively the stability of the system). Assuming $i_c= 88.72$ (grazing scenario) the outcome of the fit is nearly the same $\chi^2_{\rm red} = 1.26$, which is very close to the value  obtained when assuming coplanarity ($\chi^2=1.25$, see Section~\ref{sec:combined-rv-ttv}). It is likely that the system is very close to orbital alignment, as observed for the majority of transiting multi-planet systems, \eg , \citealt{Figueira2012,Fabrycky2014,Ballard2016,Becker2016}.

\section{Dynamical versus non-interacting orbits}\label{sec:dynamical-versus-non}
\begin{figure}[htbp]
\includegraphics[width=\linewidth]{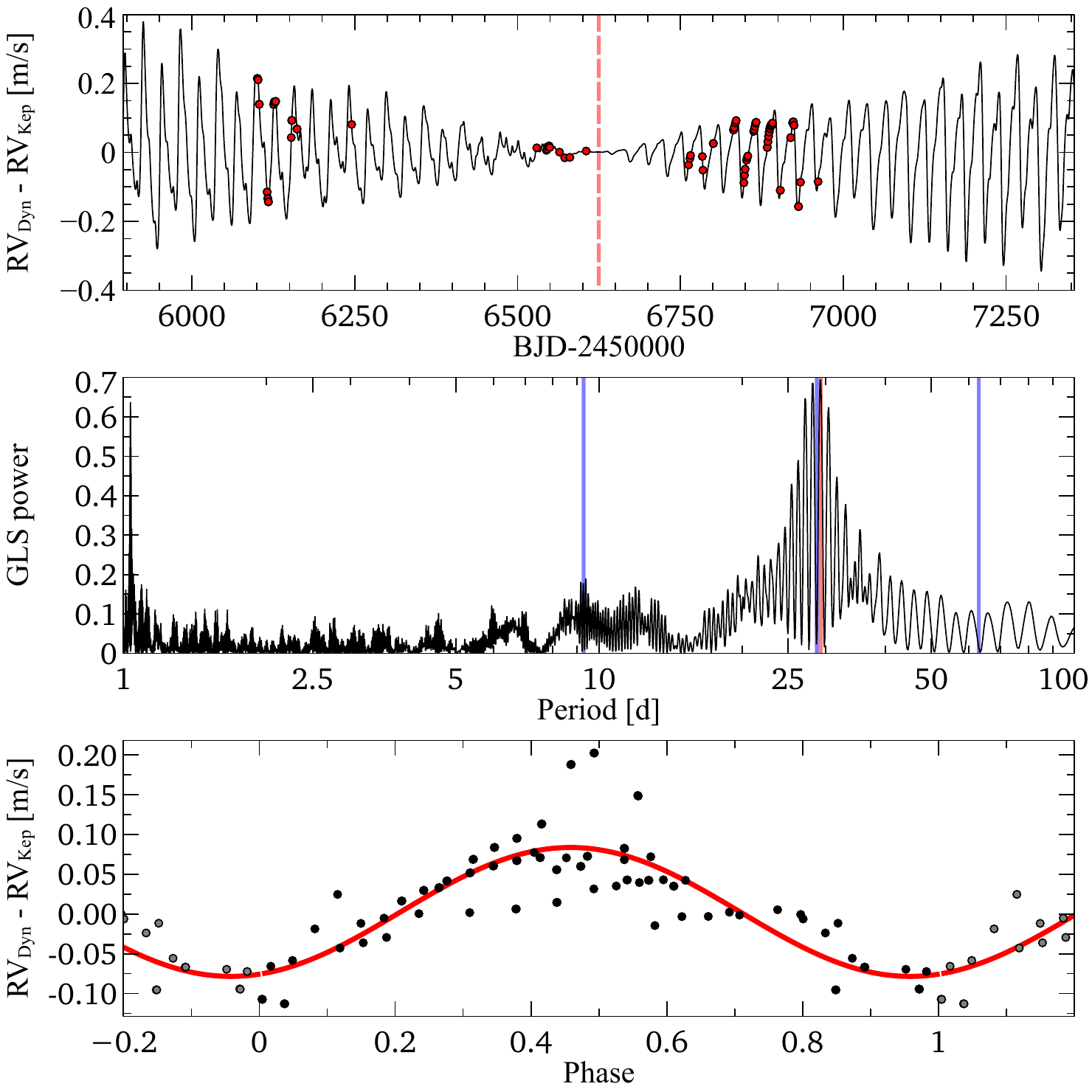}
\caption{Analysis of the RV difference ($\Delta {\rm RV}$) obtained by subtracting to the RV from dynamical simulation (${\rm RV}_{\rm Dyn}$) the RV from non-interacting Keplerian orbits (${\rm RV}_{\rm Kep}$), assuming planets with the same orbital parameters in both cases. In the upper panel, $\Delta {\rm RV}$ at the HARPS-N epochs and for a 5-year time span are shown as black dots and a black line respectively. The reference time of the orbital parameters $T_0$ is marked with a dashed red line, to emphasize the increase in $\Delta {\rm RV}$ when moving further from $T_0$. In the middle panel, the periodogram of such difference reveals the presence of a periodicity at 29.3 d (red line). As comparison, the periods of the three planets in the system are highlighted in blue. The $\Delta {\rm RV}$ phased at such period is shown in the lower panel, with the sinusoidal model marked with a red line.}
\label{fig:Dyn_vs_Kep_phased}
\end{figure}

Dynamical simulations include by definition the effects of gravitational interactions between planets. This is different from the usual approach followed in the exoplanet literature, where a series of non-interacting Keplerian orbits are used to derive the planet parameters in multiple system. While the assumption of negligible interactions between planets may hold in most of the cases, in the presence of TTVs we know that such interactions are happening. It is then worthy to analyze the differences in the RVs between the two approaches, \ie , dynamical vs. non-interacting Keplerian orbits. In order to do so, we have simulated the expected RVs of the \Ks\ system at the observational epochs of our dataset,  using the planetary parameters in Table \ref{table:pams_table_TTV} and assuming non-interacting orbits, and then we have subtracted the outcome to the dynamically-derived RVs from the same orbital parameters. The results are shown in Fig.~\ref{fig:Dyn_vs_Kep_phased}. The  ${\rm  RV_{TTV}-RV_{Kep}}$ residuals show a peak-to-peak variation of 0.30 \ms \ with a prominent periodicity at 29.3 days, \ie , very close to the orbital period of \Kc,  although these values depend strongly on the assumed orbital parameters. It is worthy to note that the difference between dynamical and Keplerian RVs diverges while moving further from the reference time of the orbital parameters $T_{\rm ref}$. The reason resides in the
small variations of the orbital parameters with time caused by planet interactions, which are implicitly taken into account in dynamical integration but ignored in the Keplerian approach.
For the \Ks\ system, after two years of observations from $T_{\rm ref}$ the peak-to-peak difference has already reached $\simeq 0.6$ \ms , \ie , a value that can negatively affect the mass determination of the planets. Assuming non-interacting planets when modeling the RVs can thus mislead the determination of the orbital parameters, \eg\ in the case of dataset spanning several years.

\section{Discussion}\label{sec:discussion}
The transiting planet \Kepb\ was previously validated by \cite{Ballard2011}. A period of 9.23 days and an upper limit of 20 \mearth\ were determined from the analysis of the light curve and high-resolution spectroscopy. From the presence of TTV in 8 {\it Kepler} quarters they deduced the presence of a second, non-transiting planet with period $\lesssim 160 $ days and mass $\lesssim 6$ \mjup .
In this paper we presented the first precise mass measurement for \Kb\ ( $8.4 \pm 1.6$ \mearth ) and the characterization of two non-transiting Neptune-mass companions, \Kc\ ($P= 28.73 \pm 0.01$ days, $M=13.1 \pm 2.7 $ \mearth ) and \Kd\ ($P= 62.9 \pm 0.2$ days, $M=20.3 \pm 3.4 $ \mearth ), obtained by simultaneously modeling TTVs and RVs through dynamical integration. 
We excluded stellar activity as a possible origin for the RV signal at $P \simeq 28$ days using all the activity diagnostics at our disposal, including the latest reduction of {\it Kepler} light curve. Nevertheless we performed standard model selection between the three-planet model and two-planet models, either with the hypothesis of the outer planet causing the TTVs and  different assumptions for the activity signal, and with the hypothesis of the perturber having period $\simeq 28$ days and no outer planets. In all cases the three-planet model resulted the favorite one with high degree of confidence. A planet in a strong MMR with \Kb\ could still produce the TTVs while having a RV semi-amplitude below our detection sensitivity, but only at the condition of assuming that the 28 signal is due to stellar activity, which is not supported by our data.

With a period ratio $P_c/P_b = 3.09$, the system is very  close to a 3:1 mean motion resonance.
The  sinusoidal shape of the TTV is then induced by the 3:1 MMR of the inner planets, with a modulation caused by the outer  planet. We performed a comparison between dynamical RVs with those calculated assuming non-interacting planets, using the orbital parameters of the \Ks\ system, and showed that, for this specific system, the difference for a dataset spanning several years is at the limit of detection with the state-of-the-art instruments used for planet search and characterization.

Our new determination of \Kb\ mass is in disagreement with the most-likely value of 1.6 \mearth\ (semi-amplitude of 0.5 \ms ) obtained by B11 while attempting a fit with only 8 Keck-HIRES RVs and assuming only one planet in the system. As a consequence of this assumption, they derived a likely RV jitter of $\simeq 4$ \ms , which is not confirmed by our analysis. In general, RV analysis performed on a small number of measurements should always be handled with extreme care, since the results could be affected by additional not-transiting planets with little influence on the TTV of the transiting planets but with significant RV semi-amplitude, such as \Kd .

\begin{figure}[htbp]
\includegraphics[width=\linewidth]{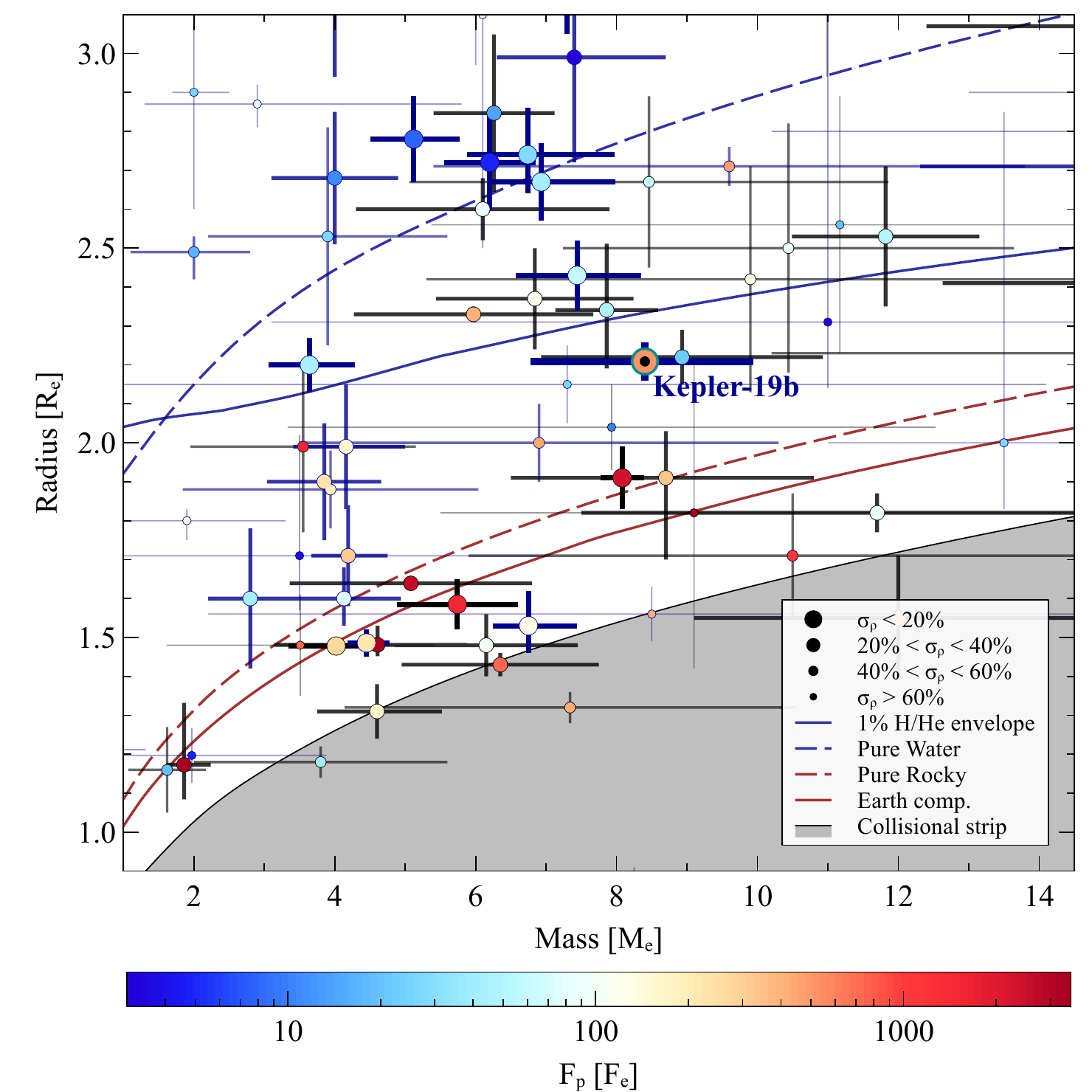}
\caption{Mass-radius for transiting exoplanets with measured masses\footnote{Automatically retrieved in February 2017 from the Nasa Exoplanet Archive, \url{http://exoplanetarchive.ipac.caltech.edu}}.
  Planets are color-coded according to the incident flux on the planet $F_p$, relative to the solar constant \fearth . Line thickness reflects the precision on density measurements. Blue error bars identify planets with TTV detection.  The dashed lines are theoretical mass-radius curves for different internal compositions, while Earth-like composition and a mixture of Earth-like core plus H/He envelope of 1\% in mass are represented by solid red and blue lines respectively \citep{Lopez2012}. The shaded gray region marks the maximum value of iron content predicted by collisional stripping \citep{Marcus2010}. \Kb\ is highlighted with a black dot.}
\label{fig:MR_Kepler19b}
\end{figure}

\Kb\ falls in the region of super-Earths with rocky cores and a significant fraction of volatiles or H/He gas, in opposition to the low-density planets characterized by TTV only (\citealt{Weiss2014,JontofHutter2016}), and well separated from the group of rocky planets with radii smaller than 2 \rearth , as can be seen in Figure \ref{fig:MR_Kepler19b}.
If we assume that the planet composition is a mixture of H/He envelope with solar composition atop a rocky core with Earth-like rock/iron abundances, we can estimate the internal structure of K19b. By employing theoretical models from \cite{LopezFortney2014}, and assuming an age of 2 Gyr, an envelope of $0.4\pm 0.3\%$ of the total mass is required to explain the observed mass and radius of \Kb . Despite the fairly high level of irradiation, its atmosphere is only moderately vulnerable to photo-evaporation due to the relatively large mass of the planet \citep{Lopez2012}. Employing the same models, we found that the mass of the primordial envelope was approximately twice ( $\simeq 1\% $) the current envelope mass.

Although the radii of the Neptune-mass planets are unknown, we can still speculate on their possible internal composition. By using the gas accretion scaling relations from \cite{LeeChiang2015},  we expect for \Kc\ and \Kd\ to have accreted significantly larger envelopes than \Kb, likely unaffected by photo-evaporation due to the larger masses of these two planets. Using Equation 22 from \cite{LeeChiang2015}, we can get a rough estimate for the size of these envelopes and find that \Kc\ should have an H/He envelope of $\simeq 4.5\%$ of its total mass and a radius of $\simeq 3.2$ \rearth , while \Kd\ should have a fraction of volatiles around 10-20\% and with a radius of $4-5$ \rearth .

Our results confirm that TTV and RV techniques can converge to planetary densities similar to the ones obtained with RV dataset only, when enough data is involved from both sides. This result support the analysis of \cite{Steffen2016}, where the discrepancy between planetary density obtained from TTV and RV noted by \cite{Weiss2014} is explained as a selection effect rather than an intrinsic problem of one of the two techniques.
Previous notable examples of agreement between RV and TTV masses are represented by the WASP-47 \citep{Becker2015,Weiss2016} and the {\it Kepler}-18 systems \citep{Cochran2011}, and the independent RV confirmation by \cite{Dai2016} of the TTV-derived masses of the planets in the K2-19 system \citep{Barros2015}. It should be noted however that \cite{2016arXiv160401265N} derived an higher RV mass for K2-19b, while the mass obtained by combining the two datasets resides halfway the two extreme values. In the case of the KOI-94 system, RV \citep{Weiss2013} and TTV \citep{Masuda2013} planetary masses agree except for planet $d$, where the TTV mass measurement is half the mass obtained by high-precision RVs.
Up to now, most of the targets characterized with TTVs are too faint for a precise, independent mass characterization with RVs. Future space-borne missions such as TESS (Transiting Exoplanet Survey Satellite, \citealt{Ricker2014}) and PLATO (PLAnetary Transits ans stellar Oscillations, \citealt{Rauer2014}) will finally shed light on this problem by providing a large number of targets bright enough for mass measurement with both techniques independently.


\begin{acknowledgements}
The HARPS-N project was funded by the Prodex Program of the Swiss Space Office (SSO), the Harvard- University Origin of Life Initiative (HUOLI), the Scottish Universities Physics Alliance (SUPA), the University of Geneva, the Smithsonian Astrophysical Observatory (SAO), and the Italian National Astrophysical Institute (INAF), University of St. Andrews, Queen's University Belfast and University of Edinburgh.

The research leading to these results received funding from the European Union Seventh Framework Programme (FP7/2007- 2013) under grant agreement number 313014 (ETAEARTH).

This work was performed in part under contract with the California Institute of Technology/Jet Propulsion Laboratory funded by NASA through the Sagan Fellowship Program executed by the NASA Exoplanet Science Institute.

AV is supported by the NSF Graduate Research Fellowship, Grant No. DGE 1144152.

This publication was made possible by a grant from the John Templeton Foundation. The opinions expressed in this publication are those of the authors and do not necessarily reflect the views of the John Templeton Foundation. This material is based upon work supported by NASA under grant No. NNX15AC90G issued through the Exoplanets Research Program.

PF acknowledges support by Funda\c{c}\~ao para a Ci\^encia e a Tecnologia (FCT) through Investigador FCT contract of reference IF/01037/2013 and POPH/FSE (EC) by FEDER funding through the program ``Programa Operacional de Factores de Competitividade - COMPETE'', and further support in the form of an exploratory project of reference IF/01037/2013CP1191/CT0001.

XD is grateful to the Society in Science$-$Branco Weiss Fellowship for its financial support.
\end{acknowledgements}


\bibliographystyle{aasjournal} 
\bibliography{Malavolta_Kepler19} 

\begin{thebibliography}{}
\expandafter\ifx\csname natexlab\endcsname\relax\def\natexlab#1{#1}\fi

\bibitem[{{Asplund} {et~al.}(2009){Asplund}, {Grevesse}, {Sauval}, \&
  {Scott}}]{Asplund2009}
{Asplund}, M., {Grevesse}, N., {Sauval}, A.~J., \& {Scott}, P. 2009, \araa, 47,
  481

\bibitem[{{Ballard} \& {Johnson}(2016)}]{Ballard2016}
{Ballard}, S., \& {Johnson}, J.~A. 2016, \apj, 816, 66

\bibitem[{{Ballard} {et~al.}(2011){Ballard}, {Fabrycky}, {Fressin},
  {Charbonneau}, {Desert}, {Torres}, {Marcy}, {Burke}, {Isaacson}, {Henze},
  {Steffen}, {Ciardi}, {Howell}, {Cochran}, {Endl}, {Bryson}, {Rowe}, {Holman},
  {Lissauer}, {Jenkins}, {Still}, {Ford}, {Christiansen}, {Middour}, {Haas},
  {Li}, {Hall}, {McCauliff}, {Batalha}, {Koch}, \& {Borucki}}]{Ballard2011}
{Ballard}, S., {Fabrycky}, D., {Fressin}, F., {et~al.} 2011, \apj, 743, 200

\bibitem[{{Baranne} {et~al.}(1996){Baranne}, {Queloz}, {Mayor}, {Adrianzyk},
  {Knispel}, {Kohler}, {Lacroix}, {Meunier}, {Rimbaud}, \& {Vin}}]{Baranne1996}
{Baranne}, A., {Queloz}, D., {Mayor}, M., {et~al.} 1996, \aaps, 119, 373

\bibitem[{{Barros} {et~al.}(2015){Barros}, {Almenara}, {Demangeon}, {Tsantaki},
  {Santerne}, {Armstrong}, {Barrado}, {Brown}, {Deleuil}, {Lillo-Box},
  {Osborn}, {Pollacco}, {Abe}, {Andre}, {Bendjoya}, {Boisse}, {Bonomo},
  {Bouchy}, {Bruno}, {Cerda}, {Courcol}, {D{\'{\i}}az}, {H{\'e}brard}, {Kirk},
  {Lachuri{\'e}}, {Lam}, {Martinez}, {McCormac}, {Moutou}, {Rajpurohit},
  {Rivet}, {Spake}, {Suarez}, {Toublanc}, \& {Walker}}]{Barros2015}
{Barros}, S.~C.~C., {Almenara}, J.~M., {Demangeon}, O., {et~al.} 2015, \mnras,
  454, 4267

\bibitem[{{Becker} \& {Adams}(2016)}]{Becker2016}
{Becker}, J.~C., \& {Adams}, F.~C. 2016, \mnras, 455, 2980

\bibitem[{{Becker} {et~al.}(2015){Becker}, {Vanderburg}, {Adams}, {Rappaport},
  \& {Schwengeler}}]{Becker2015}
{Becker}, J.~C., {Vanderburg}, A., {Adams}, F.~C., {Rappaport}, S.~A., \&
  {Schwengeler}, H.~M. 2015, \apjl, 812, L18

\bibitem[{{Benatti} {et~al.}(2016){Benatti}, {Desidera}, {Damasso},
  {Malavolta}, {Lanza}, {Biazzo}, {Bonomo}, {Claudi}, {Marzari}, {Poretti},
  {Gratton}, {Micela}, {Pagano}, {Piotto}, {Sozzetti}, {Boccato}, {Cosentino},
  {Covino}, {Maggio}, {Molinari}, {Smareglia}, {Affer}, {Andreuzzi},
  {Bignamini}, {Borsa}, {di Fabrizio}, {Esposito}, {Martinez Fiorenzano},
  {Messina}, {Giacobbe}, {Harutyunyan}, {Knapic}, {Maldonado}, {Masiero},
  {Nascimbeni}, {Pedani}, {Rainer}, {Scandariato}, \& {Silvotti}}]{Benatti2016}
{Benatti}, S., {Desidera}, S., {Damasso}, M., {et~al.} 2016, ArXiv e-prints,
  arXiv:1611.09873

\bibitem[{{Bonomo} {et~al.}(2014){Bonomo}, {Sozzetti}, {Lovis}, {Malavolta},
  {Rice}, {Buchhave}, {Sasselov}, {Cameron}, {Latham}, {Molinari}, {Pepe},
  {Udry}, {Affer}, {Charbonneau}, {Cosentino}, {Dressing}, {Dumusque},
  {Figueira}, {Fiorenzano}, {Gettel}, {Harutyunyan}, {Haywood}, {Horne},
  {Lopez-Morales}, {Mayor}, {Micela}, {Motalebi}, {Nascimbeni}, {Phillips},
  {Piotto}, {Pollacco}, {Queloz}, {S{\'e}gransan}, {Szentgyorgyi}, \&
  {Watson}}]{Bonomo2014}
{Bonomo}, A.~S., {Sozzetti}, A., {Lovis}, C., {et~al.} 2014, \aap, 572, A2

\bibitem[{{Borsato} {et~al.}(2014){Borsato}, {Marzari}, {Nascimbeni}, {Piotto},
  {Granata}, {Bedin}, \& {Malavolta}}]{Borsato2014}
{Borsato}, L., {Marzari}, F., {Nascimbeni}, V., {et~al.} 2014, \aap, 571, A38

\bibitem[{{Borucki} {et~al.}(2011){Borucki}, {Koch}, {Basri}, {Batalha},
  {Brown}, {Bryson}, {Caldwell}, {Christensen-Dalsgaard}, {Cochran}, {DeVore},
  {Dunham}, {Gautier}, {Geary}, {Gilliland}, {Gould}, {Howell}, {Jenkins},
  {Latham}, {Lissauer}, {Marcy}, {Rowe}, {Sasselov}, {Boss}, {Charbonneau},
  {Ciardi}, {Doyle}, {Dupree}, {Ford}, {Fortney}, {Holman}, {Seager},
  {Steffen}, {Tarter}, {Welsh}, {Allen}, {Buchhave}, {Christiansen}, {Clarke},
  {Das}, {D{\'e}sert}, {Endl}, {Fabrycky}, {Fressin}, {Haas}, {Horch},
  {Howard}, {Isaacson}, {Kjeldsen}, {Kolodziejczak}, {Kulesa}, {Li}, {Lucas},
  {Machalek}, {McCarthy}, {MacQueen}, {Meibom}, {Miquel}, {Prsa}, {Quinn},
  {Quintana}, {Ragozzine}, {Sherry}, {Shporer}, {Tenenbaum}, {Torres},
  {Twicken}, {Van Cleve}, {Walkowicz}, {Witteborn}, \& {Still}}]{Borucki2011}
{Borucki}, W.~J., {Koch}, D.~G., {Basri}, G., {et~al.} 2011, \apj, 736, 19

\bibitem[{{Buchhave} {et~al.}(2016){Buchhave}, {Dressing}, {Dumusque}, {Rice},
  {Vanderburg}, {Mortier}, {Lopez-Morales}, {Lopez}, {Lundkvist}, {Kjeldsen},
  {Affer}, {Bonomo}, {Charbonneau}, {Collier Cameron}, {Cosentino}, {Figueira},
  {Fiorenzano}, {Harutyunyan}, {Haywood}, {Johnson}, {Latham}, {Lovis},
  {Malavolta}, {Mayor}, {Micela}, {Molinari}, {Motalebi}, {Nascimbeni}, {Pepe},
  {Phillips}, {Piotto}, {Pollacco}, {Queloz}, {Sasselov}, {S{\'e}gransan},
  {Sozzetti}, {Udry}, \& {Watson}}]{Buchhave2016}
{Buchhave}, L.~A., {Dressing}, C.~D., {Dumusque}, X., {et~al.} 2016, \aj, 152,
  160

\bibitem[{{Burke}(2008)}]{Burke2008}
{Burke}, C.~J. 2008, \apj, 679, 1566

\bibitem[{Castelli \& Kurucz(2004)}]{Castelli2004}
Castelli, F., \& Kurucz, R.~L. 2004, arXiv, astro-ph

\bibitem[{{Cochran} {et~al.}(2011){Cochran}, {Fabrycky}, {Torres}, {Fressin},
  {D{\'e}sert}, {Ragozzine}, {Sasselov}, {Fortney}, {Rowe}, {Brugamyer},
  {Bryson}, {Carter}, {Ciardi}, {Howell}, {Steffen}, {Borucki}, {Koch}, {Winn},
  {Welsh}, {Uddin}, {Tenenbaum}, {Still}, {Seager}, {Quinn}, {Mullally},
  {Miller}, {Marcy}, {MacQueen}, {Lucas}, {Lissauer}, {Latham}, {Knutson},
  {Kinemuchi}, {Johnson}, {Jenkins}, {Isaacson}, {Howard}, {Horch}, {Holman},
  {Henze}, {Haas}, {Gilliland}, {Gautier}, {Ford}, {Fischer}, {Everett},
  {Endl}, {Demory}, {Deming}, {Charbonneau}, {Caldwell}, {Buchhave}, {Brown},
  \& {Batalha}}]{Cochran2011}
{Cochran}, W.~D., {Fabrycky}, D.~C., {Torres}, G., {et~al.} 2011, \apjs, 197, 7

\bibitem[{{Cosentino} {et~al.}(2014){Cosentino}, {Lovis}, {Pepe}, {Cameron},
  {Latham}, {Molinari}, {Udry}, {Bezawada}, {Buchschacher}, {Figueira},
  {Fleury}, {Ghedina}, {Glenday}, {Gonzalez}, {Guerra}, {Henry}, {Hughes},
  {Maire}, {Motalebi}, \& {Phillips}}]{Cosentino2014}
{Cosentino}, R., {Lovis}, C., {Pepe}, F., {et~al.} 2014, in Society of
  Photo-Optical Instrumentation Engineers (SPIE) Conference Series, Vol. 9147,
  Society of Photo-Optical Instrumentation Engineers (SPIE) Conference Series,
  8

\bibitem[{{Coughlin} {et~al.}(2016){Coughlin}, {Mullally}, {Thompson}, {Rowe},
  {Burke}, {Latham}, {Batalha}, {Ofir}, {Quarles}, {Henze}, {Wolfgang},
  {Caldwell}, {Bryson}, {Shporer}, {Catanzarite}, {Akeson}, {Barclay},
  {Borucki}, {Boyajian}, {Campbell}, {Christiansen}, {Girouard}, {Haas},
  {Howell}, {Huber}, {Jenkins}, {Li}, {Patil-Sabale}, {Quintana}, {Ramirez},
  {Seader}, {Smith}, {Tenenbaum}, {Twicken}, \& {Zamudio}}]{Coughlin2016}
{Coughlin}, J.~L., {Mullally}, F., {Thompson}, S.~E., {et~al.} 2016, \apjs,
  224, 12

\bibitem[{{Dai} {et~al.}(2016){Dai}, {Winn}, {Albrecht}, {Arriagada},
  {Bieryla}, {Butler}, {Crane}, {Hirano}, {Johnson}, {Kiilerich}, {Latham},
  {Narita}, {Nowak}, {Palle}, {Ribas}, {Rogers}, {Sanchis-Ojeda}, {Shectman},
  {Teske}, {Thompson}, {Van Eylen}, {Vanderburg}, {Wittenmyer}, \&
  {Yu}}]{Dai2016}
{Dai}, F., {Winn}, J.~N., {Albrecht}, S., {et~al.} 2016, \apj, 823, 115

\bibitem[{{Demory} {et~al.}(2016){Demory}, {Gillon}, {de Wit}, {Madhusudhan},
  {Bolmont}, {Heng}, {Kataria}, {Lewis}, {Hu}, {Krick}, {Stamenkovi{\'c}},
  {Benneke}, {Kane}, \& {Queloz}}]{Demory2016}
{Demory}, B.-O., {Gillon}, M., {de Wit}, J., {et~al.} 2016, \nat, 532, 207

\bibitem[{{Dressing} {et~al.}(2015){Dressing}, {Charbonneau}, {Dumusque},
  {Gettel}, {Pepe}, {Collier Cameron}, {Latham}, {Molinari}, {Udry}, {Affer},
  {Bonomo}, {Buchhave}, {Cosentino}, {Figueira}, {Fiorenzano}, {Harutyunyan},
  {Haywood}, {Johnson}, {Lopez-Morales}, {Lovis}, {Malavolta}, {Mayor},
  {Micela}, {Motalebi}, {Nascimbeni}, {Phillips}, {Piotto}, {Pollacco},
  {Queloz}, {Rice}, {Sasselov}, {S{\'e}gransan}, {Sozzetti}, {Szentgyorgyi}, \&
  {Watson}}]{Dressing2015}
{Dressing}, C.~D., {Charbonneau}, D., {Dumusque}, X., {et~al.} 2015, \apj, 800,
  135

\bibitem[{{Dumusque} {et~al.}(2014){Dumusque}, {Bonomo}, {Haywood},
  {Malavolta}, {S{\'e}gransan}, {Buchhave}, {Collier Cameron}, {Latham},
  {Molinari}, {Pepe}, {Udry}, {Charbonneau}, {Cosentino}, {Dressing},
  {Figueira}, {Fiorenzano}, {Gettel}, {Harutyunyan}, {Horne}, {Lopez-Morales},
  {Lovis}, {Mayor}, {Micela}, {Motalebi}, {Nascimbeni}, {Phillips}, {Piotto},
  {Pollacco}, {Queloz}, {Rice}, {Sasselov}, {Sozzetti}, {Szentgyorgyi}, \&
  {Watson}}]{Dumusque2014}
{Dumusque}, X., {Bonomo}, A.~S., {Haywood}, R.~D., {et~al.} 2014, \apj, 789,
  154

\bibitem[{{Duncan} {et~al.}(1998){Duncan}, {Levison}, \& {Lee}}]{Duncan1998}
{Duncan}, M.~J., {Levison}, H.~F., \& {Lee}, M.~H. 1998, \aj, 116, 2067

\bibitem[{{Fabrycky} {et~al.}(2014){Fabrycky}, {Lissauer}, {Ragozzine}, {Rowe},
  {Steffen}, {Agol}, {Barclay}, {Batalha}, {Borucki}, {Ciardi}, {Ford},
  {Gautier}, {Geary}, {Holman}, {Jenkins}, {Li}, {Morehead}, {Morris},
  {Shporer}, {Smith}, {Still}, \& {Van Cleve}}]{Fabrycky2014}
{Fabrycky}, D.~C., {Lissauer}, J.~J., {Ragozzine}, D., {et~al.} 2014, \apj,
  790, 146

\bibitem[{{Feigelson} \& {Babu}(2012)}]{Feigelson2012}
{Feigelson}, E.~D., \& {Babu}, G.~J. 2012, {Modern Statistical Methods for
  Astronomy}

\bibitem[{{Figueira} {et~al.}(2013){Figueira}, {Santos}, {Pepe}, {Lovis}, \&
  {Nardetto}}]{Figueira2013}
{Figueira}, P., {Santos}, N.~C., {Pepe}, F., {Lovis}, C., \& {Nardetto}, N.
  2013, \aap, 557, A93

\bibitem[{{Figueira} {et~al.}(2012){Figueira}, {Marmier}, {Bou{\'e}}, {Lovis},
  {Santos}, {Montalto}, {Udry}, {Pepe}, \& {Mayor}}]{Figueira2012}
{Figueira}, P., {Marmier}, M., {Bou{\'e}}, G., {et~al.} 2012, \aap, 541, A139

\bibitem[{{Ford}(2006)}]{Ford2006}
{Ford}, E.~B. 2006, \apj, 642, 505

\bibitem[{{Foreman-Mackey} {et~al.}(2013){Foreman-Mackey}, {Hogg}, {Lang}, \&
  {Goodman}}]{ForemanMackey2013}
{Foreman-Mackey}, D., {Hogg}, D.~W., {Lang}, D., \& {Goodman}, J. 2013, \pasp,
  125, 306

\bibitem[{{Fossati} {et~al.}(2017){Fossati}, {Marcelja}, {Staab}, {Cubillos},
  {France}, {Haswell}, {Ingrassia}, {Jenkins}, {Koskinen}, {Lanza}, {Redfield},
  {Youngblood}, \& {Pelzmann}}]{Fossati2017}
{Fossati}, L., {Marcelja}, S.~E., {Staab}, D., {et~al.} 2017, ArXiv e-prints,
  arXiv:1702.02883

\bibitem[{{Gelman} \& {Rubin}(1992)}]{Gelman1992}
{Gelman}, A., \& {Rubin}, D.~B. 1992, Statistical Science, 7, 457

\bibitem[{{Gettel} {et~al.}(2016){Gettel}, {Charbonneau}, {Dressing},
  {Buchhave}, {Dumusque}, {Vanderburg}, {Bonomo}, {Malavolta}, {Pepe}, {Collier
  Cameron}, {Latham}, {Udry}, {Marcy}, {Isaacson}, {Howard}, {Davies}, {Silva
  Aguirre}, {Kjeldsen}, {Bedding}, {Lopez}, {Affer}, {Cosentino}, {Figueira},
  {Fiorenzano}, {Harutyunyan}, {Johnson}, {Lopez-Morales}, {Lovis}, {Mayor},
  {Micela}, {Molinari}, {Motalebi}, {Phillips}, {Piotto}, {Queloz}, {Rice},
  {Sasselov}, {S{\'e}gransan}, {Sozzetti}, {Watson}, {Basu}, {Campante},
  {Christensen-Dalsgaard}, {Kawaler}, {Metcalfe}, {Handberg}, {Lund},
  {Lundkvist}, {Huber}, \& {Chaplin}}]{Gettel2016}
{Gettel}, S., {Charbonneau}, D., {Dressing}, C.~D., {et~al.} 2016, \apj, 816,
  95

\bibitem[{{Gladman}(1993)}]{Gladman1993}
{Gladman}, B. 1993, \icarus, 106, 247

\bibitem[{{Gomes da Silva} {et~al.}(2011){Gomes da Silva}, {Santos}, {Bonfils},
  {Delfosse}, {Forveille}, \& {Udry}}]{GomesDaSilva2011}
{Gomes da Silva}, J., {Santos}, N.~C., {Bonfils}, X., {et~al.} 2011, \aap, 534,
  A30

\bibitem[{{Ioannidis} {et~al.}(2016){Ioannidis}, {Huber}, \&
  {Schmitt}}]{Ioannidis2016}
{Ioannidis}, P., {Huber}, K.~F., \& {Schmitt}, J.~H.~M.~M. 2016, \aap, 585, A72

\bibitem[{{Jontof-Hutter} {et~al.}(2016){Jontof-Hutter}, {Ford}, {Rowe},
  {Lissauer}, {Fabrycky}, {Van Laerhoven}, {Agol}, {Deck}, {Holczer}, \&
  {Mazeh}}]{JontofHutter2016}
{Jontof-Hutter}, D., {Ford}, E.~B., {Rowe}, J.~F., {et~al.} 2016, \apj, 820, 39

\bibitem[{Kass \& Raftery(1995)}]{kassr95}
Kass, R.~E., \& Raftery, A.~E. 1995, Journal of the American Statistical
  Association, 90, 773

\bibitem[{{Laskar}(1993{\natexlab{a}})}]{Laskar1993a}
{Laskar}, J. 1993{\natexlab{a}}, Physica D: Nonlinear Phenomena, 67, 257

\bibitem[{{Laskar}(1993{\natexlab{b}})}]{Laskar1993b}
---. 1993{\natexlab{b}}, Celestial Mechanics and Dynamical Astronomy, 56, 191

\bibitem[{{Laskar} {et~al.}(1992){Laskar}, {Froeschlé}, \&
  {Celletti}}]{Laskar1992}
{Laskar}, J., {Froeschlé}, C., \& {Celletti}, A. 1992, Physica D: Nonlinear
  Phenomena, 56, 253

\bibitem[{{Lee} \& {Chiang}(2015)}]{LeeChiang2015}
{Lee}, E.~J., \& {Chiang}, E. 2015, \apj, 811, 41

\bibitem[{{Lopez} \& {Fortney}(2014)}]{LopezFortney2014}
{Lopez}, E.~D., \& {Fortney}, J.~J. 2014, \apj, 792, 1

\bibitem[{{Lopez} {et~al.}(2012){Lopez}, {Fortney}, \& {Miller}}]{Lopez2012}
{Lopez}, E.~D., {Fortney}, J.~J., \& {Miller}, N. 2012, \apj, 761, 59

\bibitem[{{L{\'o}pez-Morales} {et~al.}(2016){L{\'o}pez-Morales}, {Haywood},
  {Coughlin}, {Zeng}, {Buchhave}, {Giles}, {Affer}, {Bonomo}, {Charbonneau},
  {Collier Cameron}, {Consentino}, {Dressing}, {Dumusque}, {Figueira},
  {Fiorenzano}, {Harutyunyan}, {Johnson}, {Latham}, {Lopez}, {Lovis},
  {Malavolta}, {Mayor}, {Micela}, {Molinari}, {Mortier}, {Motalebi},
  {Nascimbeni}, {Pepe}, {Phillips}, {Piotto}, {Pollacco}, {Queloz}, {Rice},
  {Sasselov}, {Segransan}, {Sozzetti}, {Udry}, {Vanderburg}, \&
  {Watson}}]{LopezMorales2016}
{L{\'o}pez-Morales}, M., {Haywood}, R.~D., {Coughlin}, J.~L., {et~al.} 2016,
  \aj, 152, 204

\bibitem[{{Malavolta} {et~al.}(2016){Malavolta}, {Nascimbeni}, {Piotto},
  {Quinn}, {Borsato}, {Granata}, {Bonomo}, {Marzari}, {Bedin}, {Rainer},
  {Desidera}, {Lanza}, {Poretti}, {Sozzetti}, {White}, {Latham}, {Cunial},
  {Libralato}, {Nardiello}, {Boccato}, {Claudi}, {Cosentino}, {Covino},
  {Gratton}, {Maggio}, {Micela}, {Molinari}, {Pagano}, {Smareglia}, {Affer},
  {Andreuzzi}, {Aparicio}, {Benatti}, {Bignamini}, {Borsa}, {Damasso}, {Di
  Fabrizio}, {Harutyunyan}, {Esposito}, {Fiorenzano}, {Gandolfi}, {Giacobbe},
  {Gonz{\'a}lez Hern{\'a}ndez}, {Maldonado}, {Masiero}, {Molinaro}, {Pedani},
  \& {Scandariato}}]{Malavolta2016}
{Malavolta}, L., {Nascimbeni}, V., {Piotto}, G., {et~al.} 2016, \aap, 588, A118

\bibitem[{{Mamajek} \& {Hillenbrand}(2008)}]{Mamajek2008}
{Mamajek}, E.~E., \& {Hillenbrand}, L.~A. 2008, \apj, 687, 1264

\bibitem[{{Mandel} \& {Agol}(2002)}]{MA2002ApJ}
{Mandel}, K., \& {Agol}, E. 2002, \apjl, 580, L171

\bibitem[{{Marcus} {et~al.}(2010){Marcus}, {Sasselov}, {Hernquist}, \&
  {Stewart}}]{Marcus2010}
{Marcus}, R.~A., {Sasselov}, D., {Hernquist}, L., \& {Stewart}, S.~T. 2010,
  \apjl, 712, L73

\bibitem[{{Marzari} {et~al.}(2002){Marzari}, {Tricarico}, \&
  {Scholl}}]{Marzari2002}
{Marzari}, F., {Tricarico}, P., \& {Scholl}, H. 2002, \apj, 579, 905

\bibitem[{{Masuda} {et~al.}(2013){Masuda}, {Hirano}, {Taruya}, {Nagasawa}, \&
  {Suto}}]{Masuda2013}
{Masuda}, K., {Hirano}, T., {Taruya}, A., {Nagasawa}, M., \& {Suto}, Y. 2013,
  \apj, 778, 185

\bibitem[{{Mazeh} {et~al.}(2015){Mazeh}, {Holczer}, \& {Shporer}}]{Mazeh2015}
{Mazeh}, T., {Holczer}, T., \& {Shporer}, A. 2015, \apj, 800, 142

\bibitem[{Mor{\'e} {et~al.}(1980)Mor{\'e}, Garbow, \& Hillstrom}]{More1980}
Mor{\'e}, J.~J., Garbow, B.~S., \& Hillstrom, K.~E. 1980, User Guide for
  {MINPACK-1}, Report ANL-80-74

\bibitem[{{Mortier} \& {Collier Cameron}(2017)}]{Mortier2017}
{Mortier}, A., \& {Collier Cameron}, A. 2017, ArXiv e-prints, arXiv:1702.03885

\bibitem[{Nelder \& Mead(1965)}]{Nelder01011965}
Nelder, J.~A., \& Mead, R. 1965, The Computer Journal, 7, 308

\bibitem[{{Nespral} {et~al.}(2016){Nespral}, {Gandolfi}, {Deeg}, {Borsato},
  {Fridlund}, {Barragan}, {Grziwa}, {Korth}, {Cabrera}, {Csizmadia}, {Nowak},
  {Kuutma}, {Saario}, {Eigmuller}, {Erikson}, {Guenther}, {Hatzes}, {Montanes
  Rodriguez}, {Palle}, {Patzold}, {Prieto-Arranz}, {Rauer}, \&
  {Sebastian}}]{2016arXiv160401265N}
{Nespral}, D., {Gandolfi}, D., {Deeg}, H.~J., {et~al.} 2016, ArXiv e-prints,
  arXiv:1604.01265

\bibitem[{{Nesvorn{\'y}} {et~al.}(2012){Nesvorn{\'y}}, {Kipping}, {Buchhave},
  {Bakos}, {Hartman}, \& {Schmitt}}]{Nesvorny2012}
{Nesvorn{\'y}}, D., {Kipping}, D.~M., {Buchhave}, L.~A., {et~al.} 2012,
  Science, 336, 1133

\bibitem[{{Noyes} {et~al.}(1984){Noyes}, {Weiss}, \& {Vaughan}}]{Noyes1984}
{Noyes}, R.~W., {Weiss}, N.~O., \& {Vaughan}, A.~H. 1984, \apj, 287, 769

\bibitem[{Parviainen(2015)}]{Parviainen2015}
Parviainen, H. 2015, MNRAS, 450, 3233

\bibitem[{{Pepe} {et~al.}(2002){Pepe}, {Mayor}, {Galland}, {Naef}, {Queloz},
  {Santos}, {Udry}, \& {Burnet}}]{Pepe2002}
{Pepe}, F., {Mayor}, M., {Galland}, F., {et~al.} 2002, \aap, 388, 632

\bibitem[{{Pepe} {et~al.}(2013){Pepe}, {Cameron}, {Latham}, {Molinari}, {Udry},
  {Bonomo}, {Buchhave}, {Charbonneau}, {Cosentino}, {Dressing}, {Dumusque},
  {Figueira}, {Fiorenzano}, {Gettel}, {Harutyunyan}, {Haywood}, {Horne},
  {Lopez-Morales}, {Lovis}, {Malavolta}, {Mayor}, {Micela}, {Motalebi},
  {Nascimbeni}, {Phillips}, {Piotto}, {Pollacco}, {Queloz}, {Rice}, {Sasselov},
  {S{\'e}gransan}, {Sozzetti}, {Szentgyorgyi}, \& {Watson}}]{Pepe2013}
{Pepe}, F., {Cameron}, A.~C., {Latham}, D.~W., {et~al.} 2013, \nat, 503, 377

\bibitem[{Powell(1994)}]{Powell1994}
Powell, M. J.~D. 1994, A Direct Search Optimization Method That Models the
  Objective and Constraint Functions by Linear Interpolation, ed. S.~Gomez \&
  J.-P. Hennart (Dordrecht: Springer Netherlands), 51--67

\bibitem[{{Rauer} {et~al.}(2014){Rauer}, {Catala}, {Aerts}, {Appourchaux},
  {Benz}, {Brandeker}, {Christensen-Dalsgaard}, {Deleuil}, {Gizon}, {Goupil},
  {G{\"u}del}, {Janot-Pacheco}, {Mas-Hesse}, {Pagano}, {Piotto}, {Pollacco},
  {Santos}, {Smith}, {Su{\'a}rez}, {Szab{\'o}}, {Udry}, {Adibekyan}, {Alibert},
  {Almenara}, {Amaro-Seoane}, {Eiff}, {Asplund}, {Antonello}, {Barnes},
  {Baudin}, {Belkacem}, {Bergemann}, {Bihain}, {Birch}, {Bonfils}, {Boisse},
  {Bonomo}, {Borsa}, {Brand{\~a}o}, {Brocato}, {Brun}, {Burleigh}, {Burston},
  {Cabrera}, {Cassisi}, {Chaplin}, {Charpinet}, {Chiappini}, {Church},
  {Csizmadia}, {Cunha}, {Damasso}, {Davies}, {Deeg}, {D{\'{\i}}az}, {Dreizler},
  {Dreyer}, {Eggenberger}, {Ehrenreich}, {Eigm{\"u}ller}, {Erikson}, {Farmer},
  {Feltzing}, {de Oliveira Fialho}, {Figueira}, {Forveille}, {Fridlund},
  {Garc{\'{\i}}a}, {Giommi}, {Giuffrida}, {Godolt}, {Gomes da Silva},
  {Granzer}, {Grenfell}, {Grotsch-Noels}, {G{\"u}nther}, {Haswell}, {Hatzes},
  {H{\'e}brard}, {Hekker}, {Helled}, {Heng}, {Jenkins}, {Johansen},
  {Khodachenko}, {Kislyakova}, {Kley}, {Kolb}, {Krivova}, {Kupka}, {Lammer},
  {Lanza}, {Lebreton}, {Magrin}, {Marcos-Arenal}, {Marrese}, {Marques},
  {Martins}, {Mathis}, {Mathur}, {Messina}, {Miglio}, {Montalban}, {Montalto},
  {Monteiro}, {Moradi}, {Moravveji}, {Mordasini}, {Morel}, {Mortier},
  {Nascimbeni}, {Nelson}, {Nielsen}, {Noack}, {Norton}, {Ofir}, {Oshagh},
  {Ouazzani}, {P{\'a}pics}, {Parro}, {Petit}, {Plez}, {Poretti}, {Quirrenbach},
  {Ragazzoni}, {Raimondo}, {Rainer}, {Reese}, {Redmer}, {Reffert},
  {Rojas-Ayala}, {Roxburgh}, {Salmon}, {Santerne}, {Schneider}, {Schou},
  {Schuh}, {Schunker}, {Silva-Valio}, {Silvotti}, {Skillen}, {Snellen}, {Sohl},
  {Sousa}, {Sozzetti}, {Stello}, {Strassmeier}, {{\v S}vanda}, {Szab{\'o}},
  {Tkachenko}, {Valencia}, {Van Grootel}, {Vauclair}, {Ventura}, {Wagner},
  {Walton}, {Weingrill}, {Werner}, {Wheatley}, \& {Zwintz}}]{Rauer2014}
{Rauer}, H., {Catala}, C., {Aerts}, C., {et~al.} 2014, Experimental Astronomy,
  38, 249

\bibitem[{{Ricker} {et~al.}(2014){Ricker}, {Winn}, {Vanderspek}, {Latham},
  {Bakos}, {Bean}, {Berta-Thompson}, {Brown}, {Buchhave}, {Butler}, {Butler},
  {Chaplin}, {Charbonneau}, {Christensen-Dalsgaard}, {Clampin}, {Deming},
  {Doty}, {De Lee}, {Dressing}, {Dunham}, {Endl}, {Fressin}, {Ge}, {Henning},
  {Holman}, {Howard}, {Ida}, {Jenkins}, {Jernigan}, {Johnson}, {Kaltenegger},
  {Kawai}, {Kjeldsen}, {Laughlin}, {Levine}, {Lin}, {Lissauer}, {MacQueen},
  {Marcy}, {McCullough}, {Morton}, {Narita}, {Paegert}, {Palle}, {Pepe},
  {Pepper}, {Quirrenbach}, {Rinehart}, {Sasselov}, {Sato}, {Seager},
  {Sozzetti}, {Stassun}, {Sullivan}, {Szentgyorgyi}, {Torres}, {Udry}, \&
  {Villasenor}}]{Ricker2014}
{Ricker}, G.~R., {Winn}, J.~N., {Vanderspek}, R., {et~al.} 2014, in \procspie,
  Vol. 9143, Space Telescopes and Instrumentation 2014: Optical, Infrared, and
  Millimeter Wave, 914320

\bibitem[{{Robertson} {et~al.}(2013){Robertson}, {Endl}, {Cochran}, \&
  {Dodson-Robinson}}]{Robertson2013}
{Robertson}, P., {Endl}, M., {Cochran}, W.~D., \& {Dodson-Robinson}, S.~E.
  2013, \apj, 764, 3

\bibitem[{{Rogers}(2015)}]{Rogers2015}
{Rogers}, L.~A. 2015, \apj, 801, 41

\bibitem[{{Sneden}(1973)}]{Sneden1973}
{Sneden}, C. 1973, \apj, 184, 839

\bibitem[{{Sousa} {et~al.}(2015){Sousa}, {Santos}, {Adibekyan}, {Delgado-Mena},
  \& {Israelian}}]{Sousa2015}
{Sousa}, S.~G., {Santos}, N.~C., {Adibekyan}, V., {Delgado-Mena}, E., \&
  {Israelian}, G. 2015, \aap, 577, A67

\bibitem[{{Southworth} {et~al.}(2004){Southworth}, {Maxted}, \&
  {Smalley}}]{JKTEBOP2004}
{Southworth}, J., {Maxted}, P.~F.~L., \& {Smalley}, B. 2004, \mnras, 351, 1277

\bibitem[{{Steffen}(2016)}]{Steffen2016}
{Steffen}, J.~H. 2016, \mnras, 457, 4384

\bibitem[{{Torres} {et~al.}(2012){Torres}, {Fischer}, {Sozzetti}, {Buchhave},
  {Winn}, {Holman}, \& {Carter}}]{Torres2012}
{Torres}, G., {Fischer}, D.~A., {Sozzetti}, A., {et~al.} 2012, \apj, 757, 161

\bibitem[{{Torres} {et~al.}(2011){Torres}, {Fressin}, {Batalha}, {Borucki},
  {Brown}, {Bryson}, {Buchhave}, {Charbonneau}, {Ciardi}, {Dunham}, {Fabrycky},
  {Ford}, {Gautier}, {Gilliland}, {Holman}, {Howell}, {Isaacson}, {Jenkins},
  {Koch}, {Latham}, {Lissauer}, {Marcy}, {Monet}, {Prsa}, {Quinn}, {Ragozzine},
  {Rowe}, {Sasselov}, {Steffen}, \& {Welsh}}]{Torres2011}
{Torres}, G., {Fressin}, F., {Batalha}, N.~M., {et~al.} 2011, \apj, 727, 24

\bibitem[{{Valenti} \& {Fischer}(2005)}]{Valenti2005}
{Valenti}, J.~A., \& {Fischer}, D.~A. 2005, \apjs, 159, 141

\bibitem[{{Weiss} \& {Marcy}(2014)}]{Weiss2014}
{Weiss}, L.~M., \& {Marcy}, G.~W. 2014, \apjl, 783, L6

\bibitem[{{Weiss} {et~al.}(2013){Weiss}, {Marcy}, {Rowe}, {Howard}, {Isaacson},
  {Fortney}, {Miller}, {Demory}, {Fischer}, {Adams}, {Dupree}, {Howell},
  {Kolbl}, {Johnson}, {Horch}, {Everett}, {Fabrycky}, \& {Seager}}]{Weiss2013}
{Weiss}, L.~M., {Marcy}, G.~W., {Rowe}, J.~F., {et~al.} 2013, \apj, 768, 14

\bibitem[{{Weiss} {et~al.}(2016){Weiss}, {Deck}, {Sinukoff}, {Petigura},
  {Agol}, {Lee}, {Becker}, {Howard}, {Isaacson}, {Crossfield}, {Fulton}, \&
  {Hirsch}}]{Weiss2016}
{Weiss}, L.~M., {Deck}, K., {Sinukoff}, E., {et~al.} 2016, ArXiv e-prints,
  arXiv:1612.04856

\bibitem[{{Winn}(2010)}]{Winn2011_arXiv}
{Winn}, J.~N. 2010, ArXiv e-prints, arXiv:1001.2010

\bibitem[{{Wolfgang} \& {Lopez}(2015)}]{WolfgangLopez2015}
{Wolfgang}, A., \& {Lopez}, E. 2015, \apj, 806, 183

\bibitem[{Wright(1996)}]{Wright96a}
Wright, M.~H. 1996, in Pitman Research Notes in Mathematics, Vol. 344,
  Numerical Analysis 1995 (Proceedings of the 1995 Dundee Biennial Conference
  in Numerical Analysis), ed. D.~F. Griffiths \& G.~A. Watson ({CRC} Press),
  191--208

\bibitem[{{Zechmeister} \& {K{\"u}rster}(2009)}]{Zechmeister2009}
{Zechmeister}, M., \& {K{\"u}rster}, M. 2009, \aap, 496, 577

\end{thebibliography}

\end{document}